\documentclass[article,fleqn,usenatbib,useAMS]{pasa}%

\usepackage{graphicx}
\usepackage{amsmath}	
\usepackage{amssymb}	
\usepackage{multicol}       
\usepackage{bm}		
\usepackage{pdflscape}	
\usepackage{aas_macros}
\usepackage{hyperref} 
\usepackage[T1]{fontenc}
\usepackage{ae,aecompl}

\usepackage{xcolor}

\newcommand{\rev}[1]{{{#1}}}

\definecolor{byzantine}{rgb}{0.74, 0.2, 0.64}

\newcommand{\lsun}{\ifmmode{{\rm ~L}_\odot}\else{~L$_\odot$}\fi}
\newcommand{\sun}{\odot}
\newcommand{\Msun}{\ifmmode{{\rm ~M}_\odot}\else{~M$_\odot$}\fi}
\newcommand{\degr}{^{\circ} }
\newcommand{\sqdeg}{\,deg$^2$}
\newcommand{\arcsec}{\,arcsec}
\newcommand{\arcmin}{\,arcmin}
\newcommand{\ujybm}{\,$\mu$Jy/beam}
\newcommand{\ujy}{\,$\mu$Jy}

\title[EMU Pilot Survey]{The Evolutionary Map of the Universe Pilot Survey}

\author[Norris, Marvil, et al.]
{Ray P. Norris$^{1,2,}$
\thanks{raypnorris@gmail.com} ,
Joshua Marvil$^{2,3}$,
J. D. Collier$^{1,2,4}$,
Anna D. Kapi\'nska$^{3}$,
Andrew N. O'Brien$^{1,2,31}$,
L. Rudnick$^{5}$,
Heinz Andernach$^6$, 
Jacobo Asorey$^7$, 
Michael J. I. Brown$^8$, 
Marcus Br\"uggen$^9$, 
Evan Crawford$^1$, 
Jayanne English$^{11}$, 
Syed Faisal ur Rahman$^{10}$, 
Miroslav D. Filipovi\'c$^1$, 
Yjan Gordon$^{11}$, 
G\"ulay G\"urkan$^{12,13}$, 
Catherine Hale$^{12,30}$, 
Andrew M. Hopkins$^{14,1}$, 
Minh T. Huynh$^{12}$, 
Kim HyeongHan$^{15}$, 
M. James Jee$^{15,16}$, 
B\"arbel S. Koribalski$^{1,2}$, 
Emil Lenc$^2$, 
Kieran Luken$^{1,2}$, 
David Parkinson$^{17,18}$, 
Isabella Prandoni$^{21}$, 
Wasim Raja$^2$, 
Thomas H. Reiprich$^{19}$, 
Christopher J. Riseley$^{20,21,12}$, 
Stanislav S. Shabala$^{22}$, 
Jaimie R. Sheil$^8$, 
Tessa Vernstrom$^{12}$, 
Matthew T. Whiting$^2$, 
James R. Allison$^{26, 2}$, 
C. S. Anderson$^{2,3}$, 
Lewis Ball$^{2,27}$, 
Martin Bell$^{2,29}$, 
John Bunton$^{2}$, 
 T. J.  Galvin$^{23,12,1}$, 
 Neeraj Gupta$^{2,28}$, 
Aidan Hotan$^{12}$, 
Colin Jacka$^{2}$, 
Peter J. Macgregor$^{1,2}$, 
Elizabeth K. Mahony$^{2}$, 
Umberto Maio$^{24}$, 
Vanessa Moss$^2$, 
M. Pandey-Pommier$^{25}$, 
Maxim A. Voronkov$^2$ 
\\
\affil{$^{1}$Western Sydney University, Locked Bag 1797, Penrith, NSW 2751, Australia}
\affil{$^2$CSIRO Space \& Astronomy, P.O. Box 76, Epping, NSW 1710, Australia}
\affil{$^3$National Radio Astronomy Observatory, PO Box 0, Socorro, NM87801, USA}
\affil{$^4$The Inter-University Institute for Data Intensive Astronomy (IDIA), Department of Astronomy, University of Cape Town, Rondebosch, 7701, South Africa}
\affil{$^5$Minnesota Institute for Astrophysics, University of Minnesota, 116 Church St. SE, Minneapolis, MN 55455, USA}
\affil{$^6$Depto.\ de Astronom\'{i}a, DCNE, Universidad de Guanajuato, Cj\'on.\ de Jalisco s/n, Guanajuato, CP 36023, Mexico}
\affil{$^7$Centro de Investigaciones Energ\'eticas, Medioambientales y Tecnol\'ogicas (CIEMAT), Av. Complutense, 40, 28040 Madrid, Spain}
\affil{$^8$School of Physics and Astronomy, Monash University, Clayton, VIC 3800, Australia}
\affil{$^9$University of Hamburg, Hamburger Sternwarte, Gojenbergsweg 112, 21029 Hamburg, Germany}
\affil{$^{10}$Institute of Space and Planetary Astrophysics (ISPA), University of Karachi (UoK), Karachi, Pakistan}
\affil{$^{11}$Department of Physics and Astronomy, University of Manitoba, Winnipeg, MB R3T 2N2, Canada}
\affil{$^{12}$CSIRO Space \& Astronomy, PO Box 1130, Bentley WA 6102, Australia}
\affil{$^{13}$Thüringer Landessternwarte, Sternwarte 5, D-07778 Tautenburg, Germany}
\affil{$^{14}$Australian Astronomical Optics, Macquarie University, 105 Delhi Rd, North Ryde, NSW 2113, Australia}
\affil{$^{15}$Yonsei University, Department of Astronomy, Seoul, Republic of Korea}
\affil{$^{16}$Department of Physics, University of California, Davis, California, USA}
\affil{$^{17}$Korea Astronomy and Space Science Institute, Daejeon  34055, Korea}
\affil{$^{18}$University of Science  and Technology, Daejeon 34113, Korea}
\affil{$^{19}$Argelander Institute for Astronomy (AIfA), University of Bonn, Auf dem H\"ugel 71, 53121 Bonn, Germany}
\affil{$^{20}$Dipartimento di Fisica e Astronomia, Universit\`a degli Studi di Bologna, via P. Gobetti 93/2, 40129 Bologna, Italy}
\affil{$^{21}$INAF -- Istituto di Radioastronomia, via P. Gobetti 101, 40129 Bologna, Italy}
\affil{$^{22}$School of Natural Sciences, University of Tasmania, Private Bag 37, Hobart, TAS 7001, Australia}
\affil{$^{23}$International Centre for Radio Astronomy Research, Curtin University, Bentley, WA 6102, Australia}
\affil{$^{24}$INAF - Observatory of Trieste, via G. Tiepolo 11, 34143 Trieste, Italy}
\affil{$^{25}$University Claude Bernard Lyon 1, Bâtiment Quai 43 - 2ème étage, 28, avenue Gaston Berger, 69622 Villeurbanne Cedex,
France}
\affil{$^{26}$Sub-Dept. of Astrophysics, Department of Physics, University of Oxford, Denys Wilkinson Building, Keble Rd., Oxford, OX1 3RH, UK}
\affil{$^{27}$SKA Observatory, Jodrell Bank, Lower Withington, Macclesfield, Cheshire SK11 9FT, UK}
\affil{$^{28}$IUCAA, Post Bag 4,
Ganeshkhind, Pune University Campus,Pune 411 007, India}
\affil{$^{29}$School of Mathematical and Physical Sciences, University of Technology Sydney}
\affil{$^{30}$School of Physics and Astronomy, Institute for Astronomy, University of Edinburgh, Royal Observatory, Blackford Hill, EH9 3HJ Edinburgh, UK}
\affil{$^{31}$Department of Physics, University of Wisconsin-Milwaukee, P.O. Box 413, Milwaukee, WI 53201, USA}
}

\jid{PASA}
\doi{10.1017/pas.\the\year.xxx}
\jyear{\the\year}

\hypersetup{colorlinks,citecolor=blue,linkcolor=blue,urlcolor=blue}

\begin{document}

\begin{frontmatter}
\maketitle

\begin{abstract}
We present the data and initial results from the first Pilot Survey of the Evolutionary Map of the Universe (EMU), observed at 944 MHz with the Australian Square Kilometre Array Pathfinder (ASKAP) telescope. 
The survey covers 270 \sqdeg of an area covered by the Dark Energy Survey, reaching a depth of 25--30 \ujybm\ rms at a spatial resolution of $\sim$ 11--18 arcsec, resulting in a catalogue of $\sim$ 220,000  sources, of which $\sim$ 180,000 are single-component sources. Here we present the catalogue of single-component sources, together with (where available) optical and infrared cross-identifications, classifications, and redshifts. This survey explores a new region of parameter space compared to previous surveys.  Specifically, the EMU Pilot Survey has a high density of sources, and also a high sensitivity to low surface-brightness emission.  These properties result in the detection of types of sources that were rarely seen in or absent from previous surveys. We present some of these new results here.
\end{abstract}

\end{frontmatter}




\section{INTRODUCTION }
\label{sec:intro}

Large radio surveys provide substantial samples of galaxies for studying
cosmology. They also reveal rare but important stages of galaxy
evolution and expand the volume of observed parameter space. Before the survey described here took place, about 2.5 million radio sources were  known. That figure is about to increase  by about two orders of magnitude  \citep{norris2017}, primarily due to using innovative technology in the development of new radio telescopes and upgrading of older radio telescopes. These technological developments will enable several large radio surveys, which are expected to drive a rapid advance in knowledge. Figure 1 shows the historical growth of these surveys.

\begin{figure*}[h]
\begin{center}
\includegraphics[width=18cm,  angle=0]{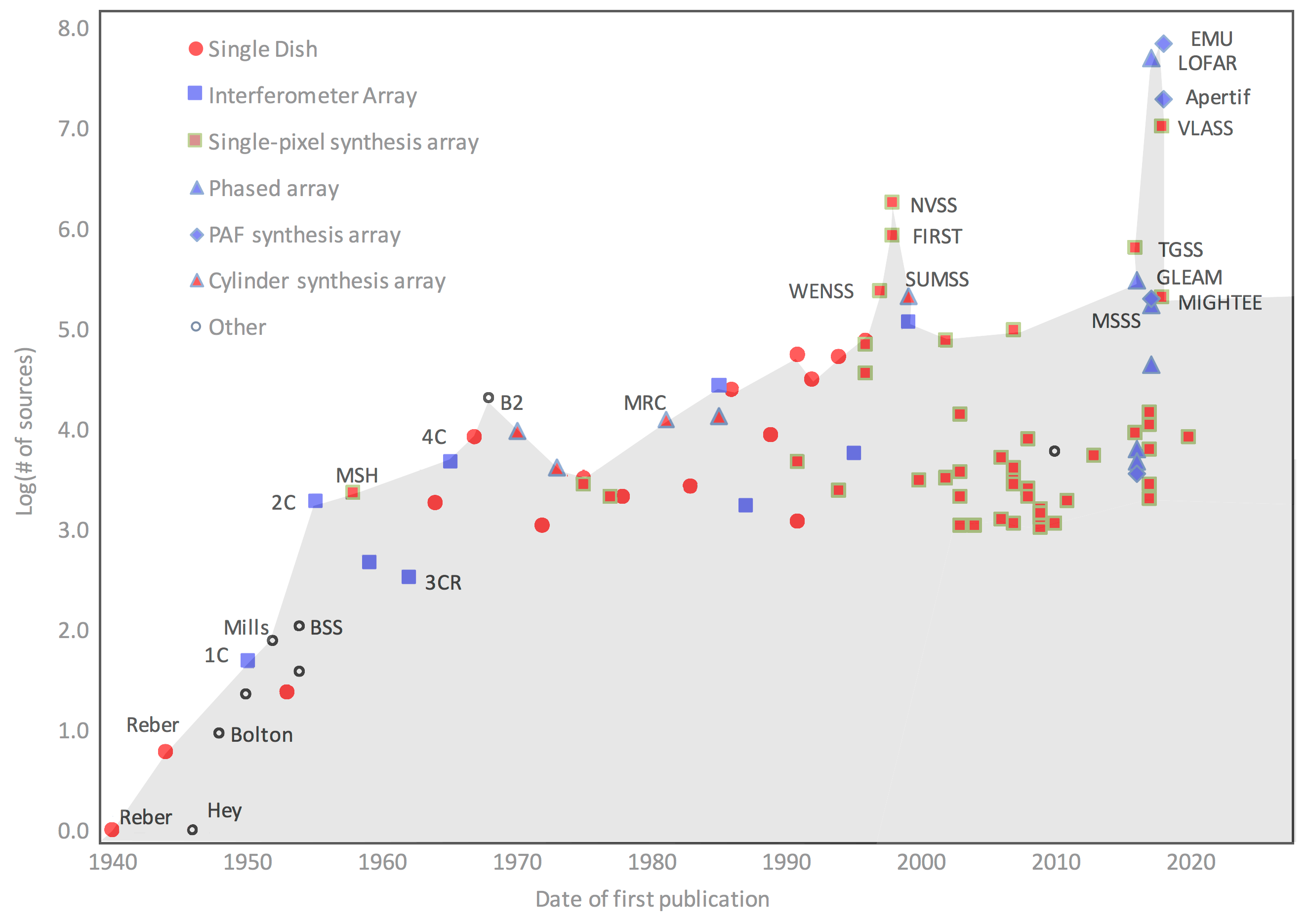}
\caption{The number of known extragalactic radio sources discovered by surveys as a
function of time, adapted from \citet{norris2017}.
The symbols
indicate the type of telescope used to make the survey, and are fully described in \citet{norris2017} . The dates and survey size are based on estimates made in 2017, and some later surveys \citep[e.g. RACS][with 2.8 million sources]{racs},  are missing from this plot. Survey abbreviations and references are given in \citet{norris2017}.
The shading under the curve is merely to improve readability.
 }
\label{srccnt}
\end{center}
\end{figure*}

\begin{figure}
\begin{center}
\includegraphics[width=8cm, angle=0]{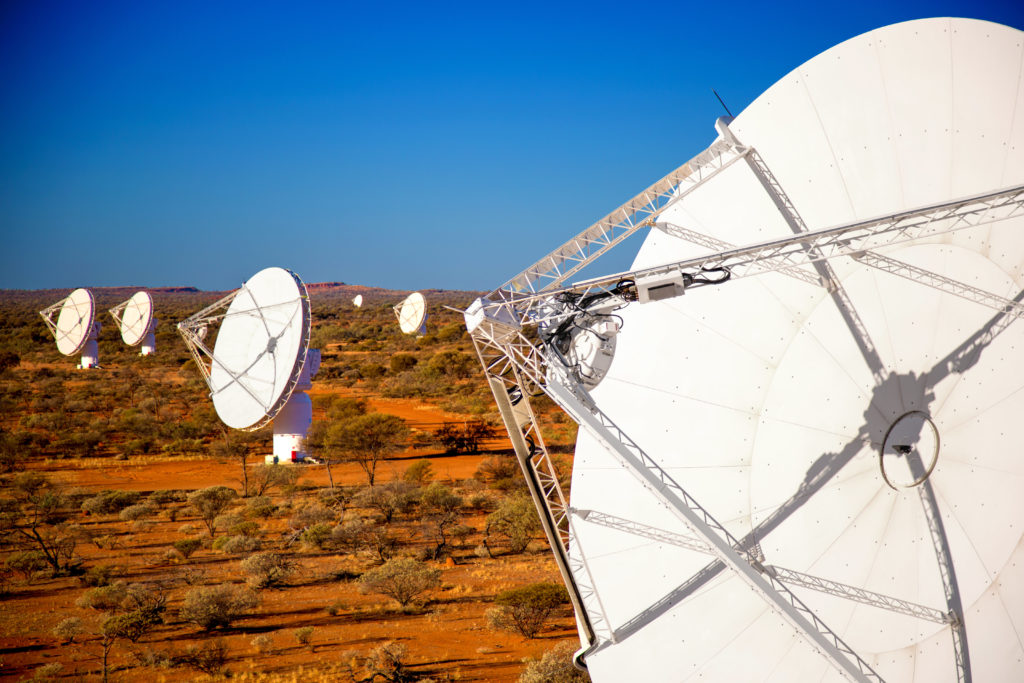}
\caption{Some of the ASKAP antennas equipped with Phased Array Feeds, located in the Murchison Region of Western Australia.  Photo credit: CSIRO
 }
\label{fig:askap}
\end{center}
\end{figure}

\begin{figure}
\begin{center}
\includegraphics[width=8cm, angle=0]{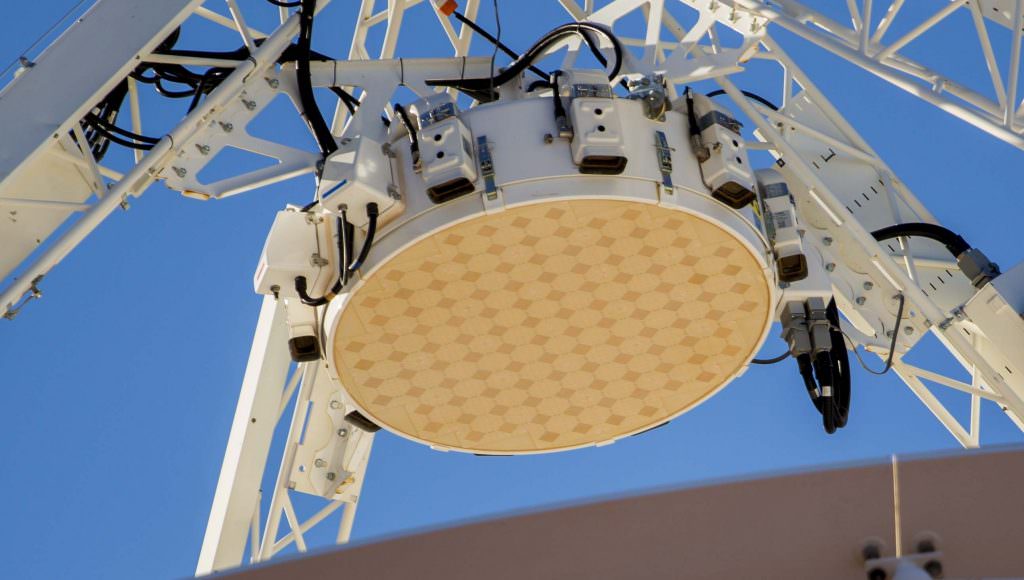}
\caption{One of the Phased Array Feeds. Each square on the chequerboard is an antenna element connected to two receivers.  Photo credit: CSIRO
 }
\label{fig:paf}
\end{center}
\end{figure}

One of these new telescopes is the Australian Square Kilometre Array Pathfinder,  \citep[ASKAP,][]{johnston07, johnston08, mcconnell16, hotan21}  which consists of 36 12-metre antennas spread over a region 6 km in diameter at the Murchison Radio Astronomy Observatory in Western Australia, shown in Figure \ref{fig:askap}. At the focus of each antenna is an innovative phased-array feed \citep[PAF:][]{PAF} of 94 dual-polarisation pixels (Figure \ref{fig:paf}). As a result, ASKAP has an instantaneous  field of view up to 30 \sqdeg, producing a much higher survey speed than that of previous synthesis arrays. The antennas are a novel 3-axis design, with the feed and reflector rotating  to ensure a constant position angle of the PAF and sidelobes on the sky.

The first all-sky survey undertaken by ASKAP was the Rapid ASKAP Continuum Survey \citep{racs} which surveyed the entire  sky south of Declination +41$\degr$ to a median rms of about 250 \ujybm. Apart from its astrophysical importance, this survey will also generate a sky-model \rev{(Hale et al., in preparation)}
to facilitate the calibration of subsequent deeper observations with ASKAP.
 
ASKAP will conduct a deep all-sky continuum survey known as the Evolutionary Map of the Universe \citep[EMU:][]{emu}. The primary goal of EMU is to make a deep (10--20 \ujybm\ rms) radio continuum survey of the entire Southern Sky, extending as far North as $+30\degr$. 
EMU is expected to generate a catalogue of as many as 70 million galaxies. 

In preparation for the full EMU survey, we conducted the EMU Pilot Survey (EMU-PS) with the  goal of testing the planned EMU survey strategy and the processing pipeline. In designing the pilot survey, we adopted the following boundary conditions:
\begin{itemize}
\item  Declination $ <  -30 \degr $ (to avoid potentially poor $u,v$ coverage near the equator).
\item Galactic latitude $ > +20 \degr $  (to avoid the strong diffuse emission in the Galactic plane).
\item Sufficiently far from the Sun to avoid solar interference, or night-time observation.
\item A single area of  240--300 \sqdeg, of 10--12 hour observations each on contiguous fields, to form a rectangular area. Cosmological analyses are optimised if the area is as square as possible. 
\item Fields overlapped by a small amount to provide uniform sensitivity.
\item Frequency band chosen to avoid any radio frequency interference and maximise survey speed, subject to constraints on resolution
and confusion.
\item Field that is well studied at other wavelengths to maximise the scientific value. 
\end{itemize} 
These  boundary conditions were satisfied by the  survey described in this paper. The survey specifications are given in Table \ref{specs}. 

\begin{table*}[t]
\centering
\caption{EMU Pilot Survey Specifications}
\begin{tabular}{ll}
\hline
Area of survey & 270 \sqdeg  \\
Synthesised beamwidth & typically 13 arcsec $\times 11$ arcsec FWHM \\
Frequency range & 800 -- 1088 MHz \\
Observing configuration & ``closepack36'' 
with pitch 0.9$\degr$ and no interleaving \\
Total integration time & 10 $\times$ 10 hours \\
\hline
\end{tabular}
\label{specs}
\end{table*}

An area of sky within the Dark Energy Survey \citep[DES:][]{DES} was chosen so that we could access the excellent optical \rev{photometric} data available from DES. EMU and DES have a Memorandum of Understanding that enables data to be shared between the two projects.

The observations were taken and processed in late 2019. It should be emphasised that, at the time of observation, commissioning of the telescope and its processing software were not yet complete, so that there are known telescope issues and processing deficiencies which were not yet addressed. As a result, the images show some artefacts, and the rms noise level is about twice as high as we expect in the final EMU survey. Nevertheless, this is still the largest radio survey ever completed at this depth, and so a great deal of valuable science results are being obtained, some of which are discussed briefly in this paper.

Section~2 of this paper describes the observations, and Section~3 describes the data reduction. Section~4 describes the ``value-added'' data processing, and Section~5 presents the results and data access. Section~6 presents some preliminary science results.

\subsection{Nomenclature and Conventions}
\begin{table*}
\centering
\caption{Cosmological parameters used in this paper and adopted for  EMU-PS}
\label{cosmoparams}
\begin{tabular}{lll}
Description & Parameter	&	value	\\
\hline
Hubble Constant & H$_0$ & 67.36 \\
Matter density & $\Omega_m$ & 0.3153\\
Cosmological constant density &$\Omega_\Lambda$ & 0.6847\\
Optical depth to reionization &$\tau$ & 0.0544\\
Physical baryon density &$\Omega_b h^2$ & 0.02237\\
Physical Cold Dark Matter density & $\Omega_{\mathrm{CDM}} h^2$ & 0.1200\\
Physical Neutrino density & $\Omega_{\nu} h^2$ & 0.000694\\
Neutrino hierarchy & \multicolumn{2}{c}{1 massive, two massless} \\
Primordial spectral index of scalar fluctuations& $n_s$ & 0.9649 \\
Amplitude of scalar fluctuations& $A_s$  & 2.055 e--9 \\
\hline
\end{tabular}
\end{table*}

Throughout this paper and in the catalogue, we use source names in the format 
\mbox{EMU PS JHHMMSS.S$-$DDMMSS} 
and we define spectral index $\alpha$  in terms of the relationship between flux density $S$ and observing  frequency $\nu$  as $S \propto \nu^{\alpha}$.


For consistency among science results derived from EMU-PS data, we encourage the use of a consistent set of cosmological parameters \rev{in papers reporting results from EMU-PS}.
Here we assume a flat $\Lambda$CDM model,  with parameter values taken from the mean posterior of the Planck 2018 cosmology,  from paper VI \citep{planck20}, using 
\rev{a combination of Planck data,}
but with no extra, non-Planck data (e.g. no \rev{Baryon Acoustic Oscillation data}).
This results in the parameter set shown in Table \ref{cosmoparams}.

\section{Observations}
\label{obs}

\begin{table*}
\begin{center}
\caption{EMU Pilot Observation Details, }
\label{observations}
\begin{tabular}{lllllll}
\hline
Date & Field name & RA & Dec. & Target  & Cal  & Number of \\
 & &(J2000) &(J2000) & SBID & SBID & antennas \\
\hline
15 Jul 2019 & EMU\_2059-51 & 21:00:00.00 & -51:07:06.4 & 9287 & 9301 & 36\\
17 Jul 2019 & EMU\_2034-60 & 20:34:17.14 & -60:19:18.2 & 9325 & 9324 & 35\\ 
18 Jul 2019 & EMU\_2042-55 & 20:42:00.00 & -55:43:29.4 & 9351 & 9350 & 35\\
24 Jul 2019 & EMU\_2115-60 & 21:15:25.71 & -60:19:18.2 & 9410 & 9409 & 35\\
25 Jul 2019 & EMU\_2132-51 & 21:32:43.64 & -51:07:06.4 & 9434 & 9428 & 34\\
26 Jul 2019 & EMU\_2027-51 & 20:27:16.36 & -51:07:06.4 & 9437 & 9436 & 36\\
27 Jul 2019 & EMU\_2118-55 & 21:18:00.00 & -55:43:29.4 & 9442 & 9441 & 36\\
02 Aug 2019 & EMU\_2156-60 & 21:56:34.29 & -60:19:18.2 & 9501 & 9500 & 36\\
03 Oct 2019 & EMU\_2154-55 & 21:54:00.00 & -55:43:29.4 & 10083 & 10082 & 35\\
24 Nov 2019 & EMU\_2205-51 & 22:05:27.27 & -51:07:06.4 & 10635 & 10634 & 34\\
\hline
\end{tabular}
\medskip\\
 The position shown is the antenna pointing centre, corresponding to position (0,0) in Figure \ref{closepack36}. Columns 5 and 6 show the  ASKAP scheduling block identification (SBID) number for the target and calibrator observations.
\end{center}
\end{table*}

\begin{table*}
\begin{center}
\caption{EMU Pilot Processing parameters. \rev{The first column shows the parameter name used by ASKAPsoft.}}
\label{parset}
\begin{scriptsize}
\begin{tabular}{lll}
\hline
Parset name & Explanation & Value  \\
\hline
DO\_SPECTRAL\_IMAGING & Only continuum (i.e. 288 1-MHz channels) &  false \\
GRIDDER\_NWPLANES & Number of w planes &  557 \\
CLEAN\_SCALES & Scales used by multiscale CLEAN, in pixels & [0,6,15,30,45,60] \\
CLEAN\_NUM\_MAJORCYCLES & Number of major cycles of CLEAN &  [5,15] \\
CLEAN\_THRESHOLD\_MINORCYCLE & Stop CLEAN minor cycles at this threshold &  [30\%, 0.25mJy, 0.03mJy] \\
CLEAN\_MINORCYCLE\_NITER & No. of iterations of CLEAN in each minor cycle &  [400,3000] \\
SELFCAL\_METHOD & Selfcal model &  CleanModel \\
SELFCAL\_INTERVAL & Selfcal solution interval (s) in each selfcal iteration & [60] \\
PRECONDITIONER\_WIENER\_ROBUSTNESS & Briggs robustness parameter in preconditioning & 0.0 \\
RESTORE\_PRECONDITIONER\_LIST & Use a Wiener filter for the main image &  [Wiener,GaussianTaper] \\
&and a Gaussian taper for the alt image  & \\
RESTORE\_PRECONDITIONER\_WIENER\_ROBUSTNESS & Both the main and the alt image have robustness $=$ 0 & 0.0 \\
RESTORE\_PRECONDITIONER\_GAUSS\_TAPER & Taper the ``alt'' image with a $30 \times 30$ arcsec Gaussian &  [30arcsec, 30arcsec, 0$\degr$] \\
LINMOS\_CUTOFF & Cut off the beams at 0.2 of peak when mosaicing them & 0.2 \\
DO\_CONTCUBE\_IMAGING & Image using a ``Continuum cube''?&  true \\
\hline
\end{tabular}
\end{scriptsize}
\medskip\\
\end{center}
\end{table*}

ASKAP has 36 antennas, all but six of which are within a region of 2.3km diameter, with the outer 6 extending the baselines up to 6.4 km. In all Pilot Survey observations, as many of the 36 antennas were used as possible. However, in some cases a few antennas were omitted because of maintenance or hardware issues. The actual number of antennas used  is shown in Table \ref{observations}.

At the prime focus of each antenna is a phased array feed (PAF), which subtends a solid angle of about 30 \sqdeg\ of the sky. The PAF consists of 188 single-polarisation dipole receivers. A weighted sum of the outputs of  groups of these receivers is used to form 36 dual-polarisation ``beams''. Individual dipole receivers will, in general, contribute to more than one beam, so that adjacent beams are not completely independent. The 36 beams together cover an area of about 30 \sqdeg\ on the sky, which we refer to as a ``tile''. 
 
 \begin{figure}
\begin{center}
\includegraphics[width=7cm]{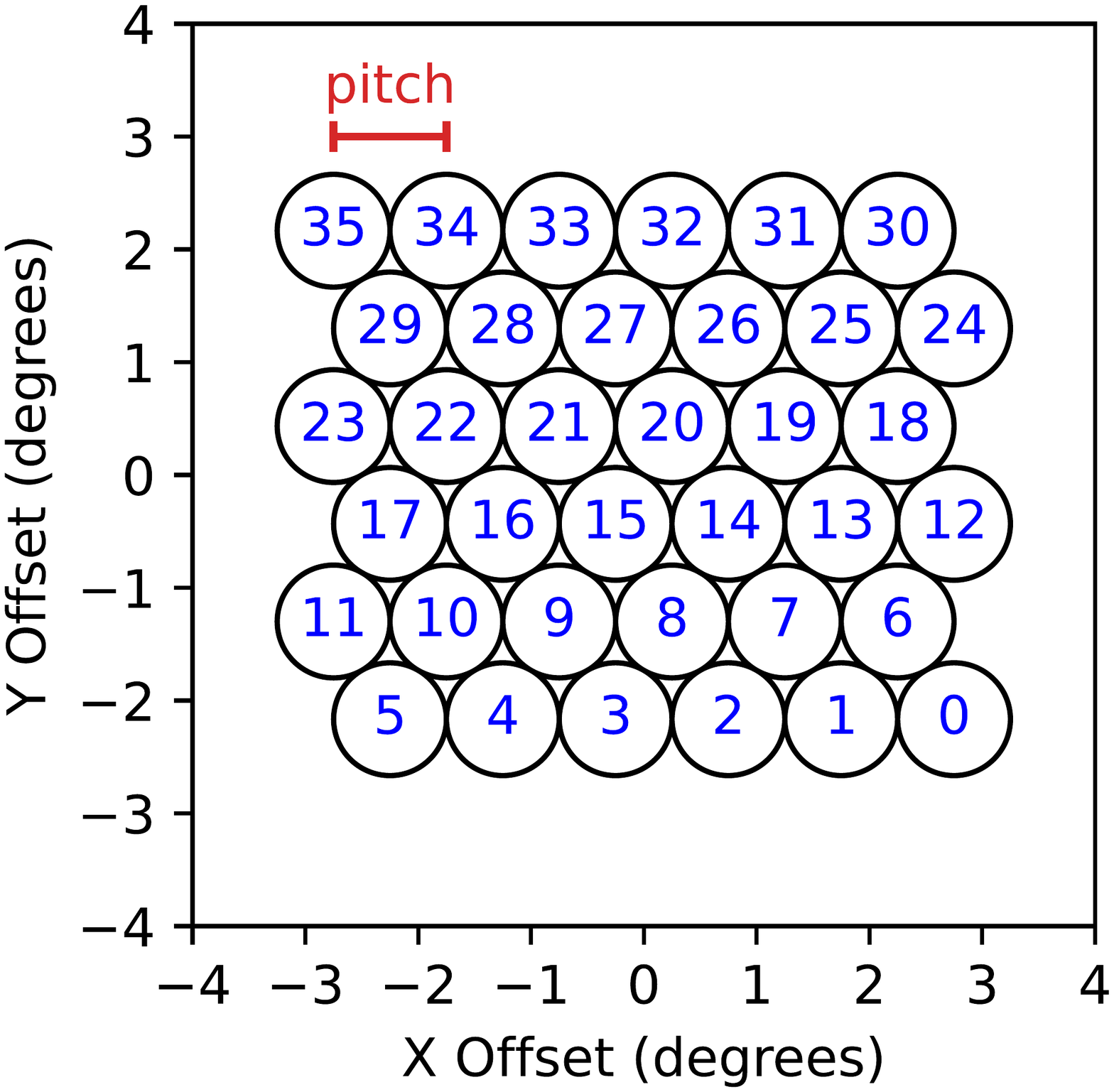}
\caption{The arrangement of the 36 ASKAP beams in the ``closepack36'' configuration. The beams are numbered from 0 to 35 \citep[diagram adapted from][]{mcconnell19}. The circles shown are for illustration only. For EMU-PS, the actual full width half maximum of each beam is $\sim 1.5\degr$ at the band centre, and the pitch spacing is $0.9\degr$, giving an approximately uniform sensitivity over the field of view.}
 
\label{closepack36}
\end{center}
\end{figure}
 \begin{figure}
\centering
\includegraphics[width=5cm, angle=0]{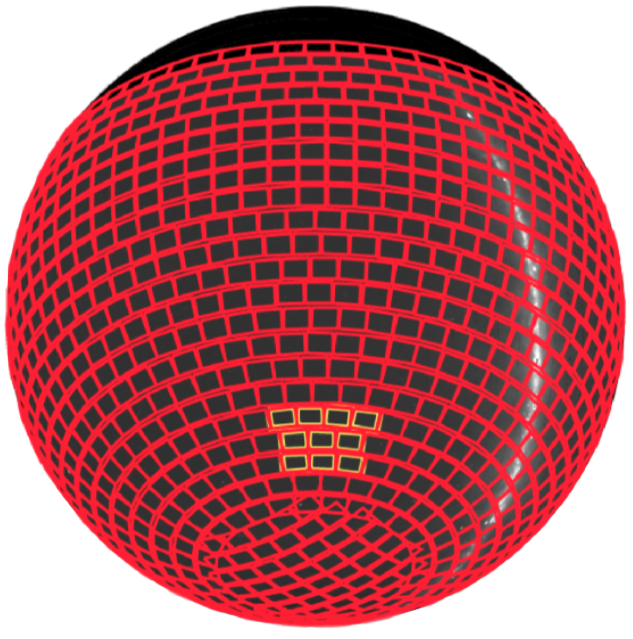}
\caption{The sky tiling scheme adopted for the EMU-PS. The red rectangles covering the celestial sphere show the tiles planned for the EMU survey, and the orange area indicates the ten tiles of the EMU-PS. The white strip shows the Galactic plane, and the south celestial pole is at the bottom of the figure.}
\label{tiles}
\end{figure} 
There are several ways of arranging the individual beams within the tile. For EMU-PS, we use a hexagonal arrangement of the 36 beams with six rows of six beams, known as “closepack36”  \citep{hotan21}, shown in Figure \ref{closepack36}. 
This configuration provides  more uniform coverage than the widely used rectangular array known as ``square\_6x6''. The spacing between the beams is known as the ``pitch'', and is set to 0.9$\degr$ for EMU-PS. In some other ASKAP observations, interleaved observations are taken, with the antenna pointing position shifted by half the pitch, to provide better uniformity. However, this is not necessary for the EMU-PS because of the combination of our \rev{lower} observing frequency and the closepack36 configuration.

The weights of the individual beams are initially calibrated by observing the Sun, placed successively at the centre of each beam, and then adjusting the weights for maximum signal-to-noise. A radiator at the vertex of each antenna (the On-Dish Calibrator, or ODC) enables the gain of each receiver to be monitored, and the weight solution initially obtained from solar observations may be  updated if necessary using these ODC measurements.

Before (or sometimes after) the observation of each target, the calibrator source PKS\,1934--638 is observed for 200\,s at the centre of each of the 36 beams to provide bandpass and gain calibration. This calibration observation takes about 2 hours. 

The positions of the tiles are chosen using a tiling scheme which will be used for the main EMU survey, shown in Figure \ref{tiles}. At most declinations, the tiles are aligned with lines of constant declination. At the south polar cap (below declination -71.81$\degr$), they are arranged in a rectangular grid as shown in Figure \ref{tiles}. Using this scheme, the sky south of Declination +30$\degr$ is covered by 1280 tiles. Overlaps between tiles amount to less than 5\% of the total area covered.

The pilot survey was  observed with ASKAP in the period 15 July to  24 November 2019. In some cases,  the initial observations were subsequently found to be faulty, in which case the field was re-observed. Table \ref{observations} shows the details of the observations that were used in the final data product. 

\begin{figure}
\begin{center}
\includegraphics[width=8cm, angle=0]{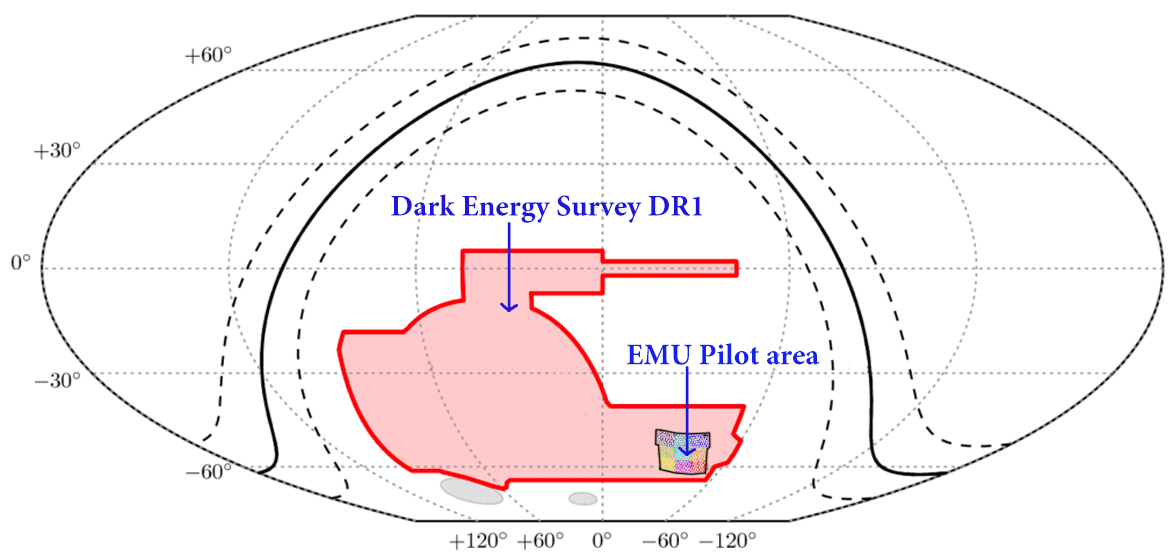}
\caption{The location of the EMU Pilot Survey area on the sky within DES DR1, adapted from \citet{DES}. \rev{The diagram is in equatorial coordinates, and the solid line marks the Galactic plane, flanked by two dashed lines showing Galactic latitude $\pm 10 \degr$}.
 }
\label{DESlocation}
\end{center}
\end{figure}

\begin{figure}
\begin{center}
\includegraphics[width=8cm, angle=0]{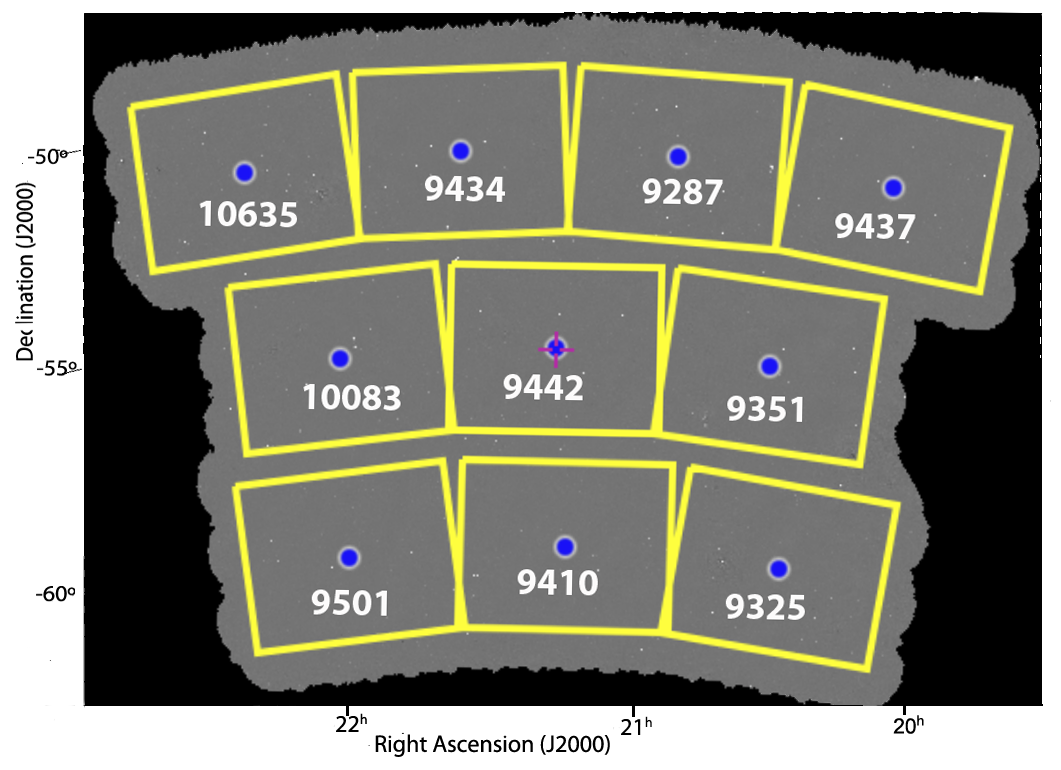}
\caption{The arrangement of the ten individual ASKAP tiles on the sky for EMU-PS with their SBID numbers as listed in Table \ref{observations}.The rectangles are separated in this diagram for clarity, but there is actually overlapping coverage as illustrated by the greyscale background.
 }
\label{pointings}
\end{center}
\end{figure}

The survey consists of a  10 hour observation of each of the ten tiles, each accompanied by a  calibration observation as described above. No further calibration
is performed during the observation. The location of the survey area is shown in Figure \ref{DESlocation}, and the details of the pointing centres are shown in Table \ref{observations} and in Figure \ref{pointings}.

\section{Pipeline Data Reduction}

We process the data using the ASKAPsoft pipeline \citep{whiting17,whiting20, guzman19} with the parameters shown in Table \ref{parset}, \rev{and using 2 arcsec square pixels. All parameter names, shown in italics in this section, are included in Table \ref{parset}.}  

The ASKAP correlator generates 16384 spectral-line channels 
and, for EMU data, we start by  averaging these to 288 1-MHz channels to reduce the computational load \rev{(i.e. \small{$DO\_SPECTRAL\_IMAGING = false$ and $DO\_CONTCUBE\_IMAGING = true$)}}

Weighting and tapering in ASKAPsoft are done using a Wiener filter preconditioning technique, which is computationally more efficient than traditional tapering and weighting. \rev{(i.e. \small{$ RESTORE\_PRECONDITIONER\_LIST = [Wiener,GaussianTaper] $}) for the main and alt image respectively }

To choose the robustness \citep{briggs95}, we conducted tests on part of the GAMA23 field (at about Right Ascension 23:00, Declination $-$32:00; \citet{leahy19}, Prandoni et al., in preparation), in both Stokes I (total intensity) and Stokes V (circular polarisation) resulting in the values shown in Table \ref{robustness}.  At lower (more negative) values of robustness the rms increases because the near-uniform weighting discards information. At higher (more positive) values of robustness, corresponding to near-natural weighting, the V rms continues to decrease but the I rms increases presumably because of (a) confusion, (b) poorer $u,v$ coverage leading to increased sidelobes, (c) increased radio-frequency interference on short baselines. Based on these results, we choose a robustness of 0.0 as an optimum value for the EMU pilot survey. \rev{i.e. \small{$PRECONDITIONER\_WIENER\_ROBUSTNESS=0.0$}}. Although robustness +0.5 has a slightly lower rms, it has a significantly increased beam size. No further tapering is used in the main image.

\begin{table}
\begin{center}
\caption{Results of tests to measure the optimum robustness. 
}
\label{robustness}
{\small
\begin{tabular}{lllll}
\hline
Briggs	&	bmaj	&	bmin	&	rms (I)	&	rms (V)	\\
robustness	&	(arcsec)	&	(arcsec)	&	\ujybm	&	\ujybm	\\
\hline
2	&	25	&	21	&	46	&	11	\\
1	&	23	&	19	&	29	&	12	\\
0.5	&	17	&	14	&	20	&	13	\\
0	&	13	&	11	&	21	&	16	\\
$-$0.5	&	9	&	9	&	33	&	29	\\
$-$1	&	8	&	7	&	56	&	48	\\
$-$2	&	8	&	7	&	73	&	61	\\
\hline
\end{tabular}}
\end{center}
Tests were conducted at 888 MHz, using a 10-hour ASKAP observation on arrays of 33 (for Stokes V, using SB8129 on a field close to UV Ceti) and 35 (for Stokes I, using SB8137 on the GAMA23 field) antennas. Columns 2 and 3 given the major and minor axes of the restoring beam, and columns 4 and 5 gives the measured rms values in (a) a source-free region of the Stokes I image, and (b) the Stokes V image, which is almost source-free. The results have been scaled to a 10-hour observation on an array of 36 antennas.
\medskip\\
\end{table}

The non-coplanarity of ASKAP is managed using the w-projection technique \citep{cornwell08, rau09}, using a total of 557 w-planes \rev{(i.e. \small{$GRIDDER\_NWPLANES=557$}}).  The data are gridded using multi-frequency synthesis, and deconvolved using a multi-frequency multi-scale CLEAN, using \rev{\small{$CLEAN\_SCALES$ of [0,6,15,30,45,60] pixels}, which} gives 6 scales up to ten times the clean beam size. After initial imaging and cleaning (using 5 major cycles: \rev{\small{$CLEAN\_NUM\_MAJOR\_CYCLES$}}, with 400 iterations in each minor cycle, down to a limit of 0.25 mJy), the data are  given one iteration of phase selfcal using the  output of the previous CLEAN \rev{(i.e. \small{$SELFCAL\_METHOD=CLEAN$)}} before the final imaging and cleaning (15 major cycles with up up to 3000 iterations in each minor cycle, with minor cycles triggering a major cycle when they reach a 30\% CLEAN limit, to a clean limit of 30 \ujy: \rev{ i.e. \small{$CLEAN\_THRESHOLD\_MINORCYCLE$)}}.  Two images are produced by the pipeline: the main image at full resolution, and an alternative (``alt'') image tapered to a 30-arcsec resolution, which is optimised for faint diffuse emission. The alt image is not used in this paper.

The multi-frequency synthesis imaging uses a Taylor term technique \citep{rau11} over the 288 MHz bandwidth to account for the spectral variation of each source.  We use two terms in the Taylor expansion, resulting in two planes called TT0 and TT1. The TT0 plane is the zeroth-order term, corresponding to the total intensity of each pixel integrated over the full bandwidth. TT1 is the first-order term and allows the spectral indices at each pixel to be measured as $\alpha$ = TT1/TT0.  

Primary beam correction is applied to each beam, \rev{and beams are combined using a weighted mean down to a cutoff of 20\% of the peak (i.e. \small{$LINMOSCUTOFF =0.2$}}) assuming a Gaussian primary beam shape. Future ASKAP surveys will use a beam shape based on holographic measurements, but that was not available for EMU-PS. Using the Gaussian beam approximation increases calibration errors and the rms noise level.

Source extraction uses the ``{\it Selavy}'' software tool \citep{whiting12,whiting17} which  identifies ``islands'' of emission higher than three times the local rms in the image, using a flood-fill technique, and then fits Gaussian components to peaks of emission within the islands. Only components and islands greater than five times the local rms are retained.


In the EMU \rev{initial} public data release \rev{(defined in Section \ref{public})}, spectral indices for individual components are measured as $\alpha$ = TT1/TT0, where TT0 and TT1  are a weighted mean of the Taylor terms over the area of the component, down to a level of 5 times the local rms noise. \rev{However, this technique has been found to be unsatisfactory,  so the spectral indices in the \rev{initial} public data release should be regarded as unreliable. Our alternative technique is discussed below in Section \ref{spindex1}}


As a final step within the pipeline, the data from each scheduling block is uploaded to the data archive (but not yet released) and passed through the ASKAP continuum validation package\footnote{\url{https://confluence.csiro.au/display/askapsst/Continuum+Validation}}
using default parameters.
 This package takes the final image, noise map and {\it Selavy} catalogue as input, and produces metrics and data quality flags based on a number of validation tests\footnote{\url{https://confluence.csiro.au/display/askapsst/Continuum+validation+metrics}}. The following metrics are used for each of these tests using ASKAP data only:

\begin{itemize}
    \item Fraction of sources considered resolved, given by the difference in the integrated ($S_{\rm int}$) and peak ($S_{\rm peak}$) flux densities, and local noise $\sigma$. We consider a source to be resolved   when ($S_{\rm int} - S_{\rm peak}) / \sqrt{dS_{\rm int}^2 + dS_{\rm peak}^2 + \sigma^2} > 3$, where $dS_{\rm int}$ and $dS_{\rm peak}$ are the estimated measurement errors in $S_{\rm int}$ and $S_{\rm peak}$ 
    \item Reduced $\chi^2$ of differential Euclidean source counts
    \item Median RMS value (from {\it Selavy} noise map)
    \item Median in-band spectral index (from {\it Selavy} catalogue, measured from Taylor term images)
\end{itemize}

The following additional metrics are used with respect to selected point sources, cross-matched to the reference catalogue that provided the most matches, which for the pilot, is SUMSS \citep{sumss}, which has a resolution of $\sim$ 45 arcsec at an observing frequency of 843 MHz:

\begin{itemize}
    \item \rev{Median absolute deviation (MAD) of the  ratio of the flux density of the reference catalogue to the ASKAP flux density, after correcting for the frequency difference  assuming $\alpha = -0.8$,}
    \item Flux density ratio uncertainty, calculated from the median absolute deviation (MAD)
    \item Positional offset, given by the median compared to reference catalogue
    \item Positional offset uncertainty, calculated from the MAD
\end{itemize}

\begin{figure}
\begin{center}
\includegraphics[width=8cm]{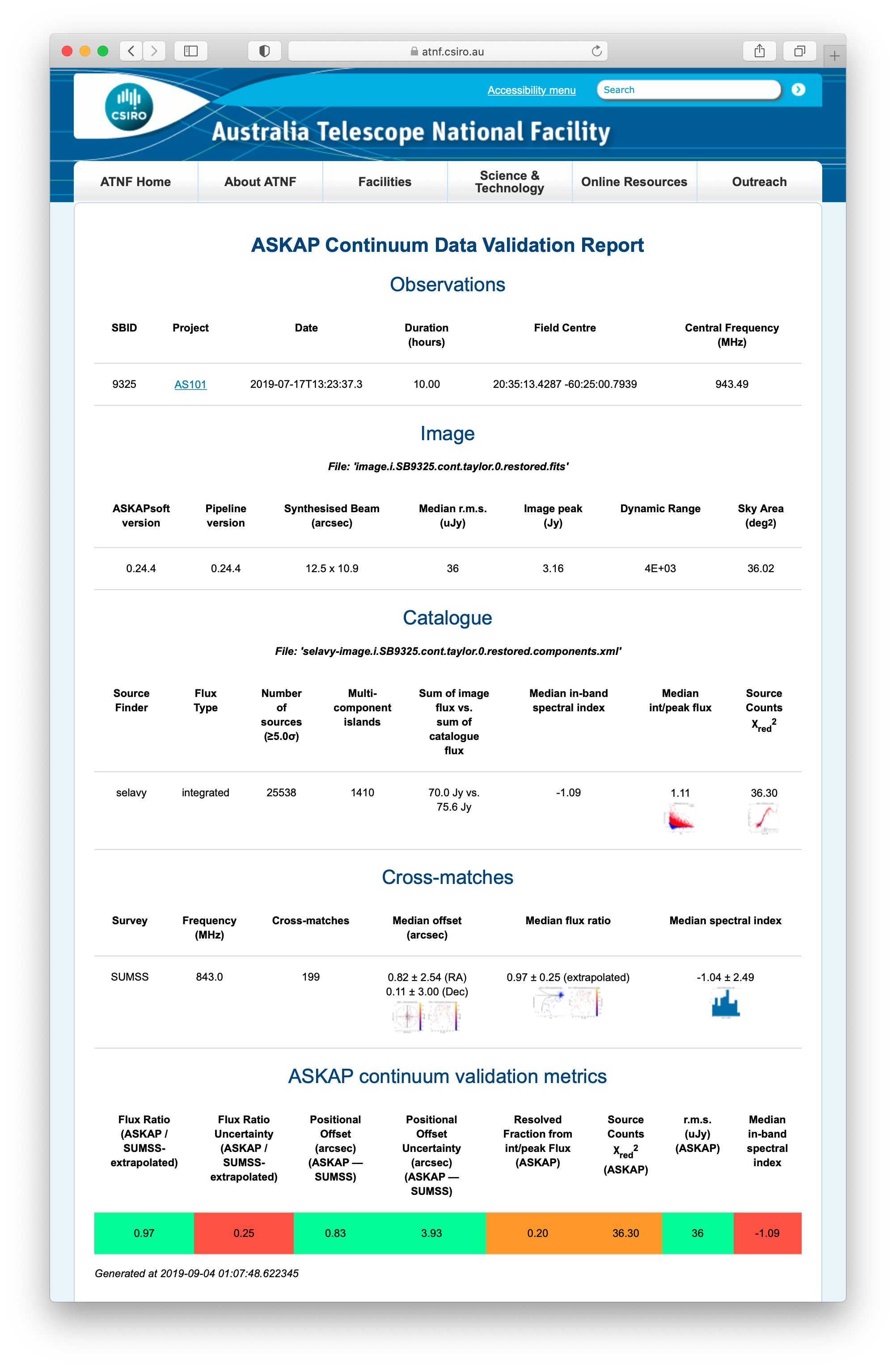}
\caption{An example validation report 
for one of the processing runs for SB9325, including the metrics and their flags. A higher-resolution version is available online$^{2}$.
}
\label{validation_report}
\end{center}
\end{figure}

An example report\footnote{
         Reports are available from     \url{https://www.atnf.csiro.au/research/ASKAP/ASKAP-validation/commissioning/AS101/SB9325/validation_image.i.SB9325.cont.taylor.0.restored__askapops_2019-09-03-101300/}
     } 
for one of the processing runs for SB9325 is shown in Figure~\ref{validation_report}, including a summary of the metrics and their flags. Each metric is flagged as good, bad or uncertain based on selected tolerance values\footnote{The metrics are described in detail in \url{https://confluence.csiro.au/display/askapsst/Continuum+validation+metrics}}. The metrics and flags are associated and archived with the data, and the validation reports are automatically uploaded as a report under project AS101\footnote{\url{https://www.atnf.csiro.au/research/ASKAP/ASKAP-validation/commissioning/AS101/}}. 

The final validation process is done by members of the EMU team and includes:
\begin{itemize}
    \item inspecting each of the validation reports described above,
    \item inspecting the images to search for artefacts,
    \item examining quantities such as the variation of restoring beam among the 36 beams used in the mosaic.
\end{itemize}

Data deemed to be acceptable are then released to the public domain on the data archive, described in Section \ref{public}. If the data are not found to be acceptable, then the data are removed from the archive and we request a re-observation.

\section{Value-Added Processing}
\label{value-added}

To mitigate some of the data issues in the \rev{initial} public data release, and to produce a unified image and source catalogue covering the full EMU-PS field, we conduct value-added processing on the \rev{initial} public data release to generate a value-added data release. This value-added processing also includes some optical and infrared ancillary data, as described below.

\subsection{Merging Tiles}
The \rev{initial} public release of the survey data consists of ten overlapping tiles, each with its own source catalogue. Simply merging these catalogues generates a large number of duplicate sources, which must be reconciled to maximise the information integrity and consistency. This approach also fails to take full advantage of the additional information, such as increased sensitivity, available where tiles overlap.

\begin{figure*}
\begin{center}
\includegraphics[width=16cm, angle=0]{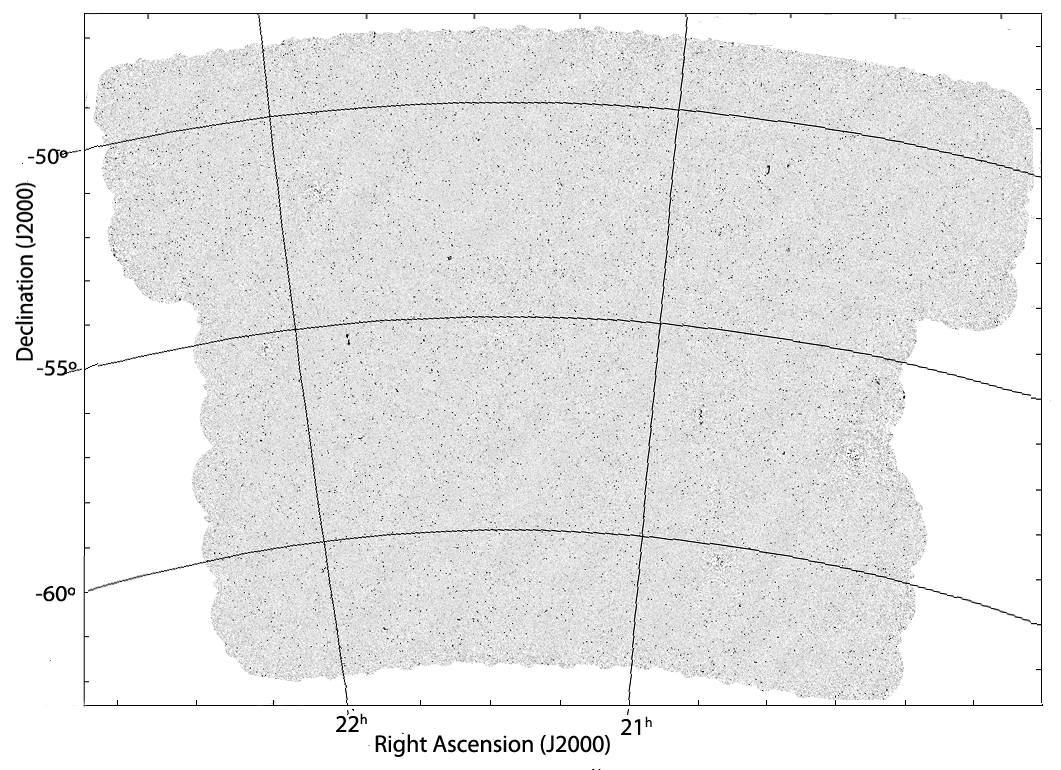}
\caption{The resulting native resolution ($13'' \times 11''$) image of the 270 \sqdeg EMU Pilot Survey, containing about 220,000 radio sources. The rms noise level is 25-30 \ujybm, and the peak flux density is 3.14 Jy/beam.}
\label{merged}
\end{center}
\end{figure*}

To overcome these issues, we  merged the ten tiles in the image plane using the ASKAPsoft task {\it linmos}, which performs a weighted average of the data in overlapping regions. The merged data set is shown in Figure \ref{merged}. We refer to this image, which has a typical spatial resolution of 11--13 arcsec,  as the ``native resolution image''.

\subsection{Convolution to a Common Restoring Beam Size}
A problem with the native resolution image is that the point spread function (psf) varies from beam to beam over the field, so that both the flux density scale, 
and also the spectral indices,  vary from beam to beam.
To overcome this problem, we created a version of the data in which each PAF beam is individually convolved with a Gaussian kernel to obtain a common circular restoring beam of 18 arcsec FWHM. We then recombined all beams into a weighted average using the ASKAPsoft task {\it linmos}. We refer to this data set as the ``convolved image''.

We then ran the {\it Selavy} source finder on the convolved image, to produce a catalogue of components and islands. This convolved catalogue has 220,102 components. These ``convolved'' data are recommended over the ``native'' data product  for measurements of flux density and spectral index. However, the data products in native resolution are still optimum for studies of morphology, or when the higher resolution is needed.

\subsection{Separation of sources into simple and complex}
 Many value-added operations, such as measuring spectral index, and cross-identifying to optical/IR catalogues, are far more complex for extended or complex sources than for simple, compact sources.  These techniques are still under development for the full EMU survey.
 
We therefore divided the source catalogue into ``simple'' and ``complex'' sources. A sophisticated technique for this separation is still under development, so for the purposes of this paper we used a simple technique in which we defined islands with only one component (specifically, with {\it has\_siblings $>$ 0 })  to be simple, and all other islands are defined to be ``complex''. This technique results in a catalogue of 178,921 components, so that about 81\% of sources in the catalogue are ``simple''. 

 \rev{Many ``complex'' sources are classical FRI or FR\,II sources \citep{FR}, but our high sensitivity to low surface brightness  has also enabled the detection of several peculiar-looking sources that are quite unlike those seen in earlier surveys such as NVSS \citep{nvss} or FIRST \citep{white97}. In  Section \ref{science}, we discuss a small sample of these peculiar objects, which will be further explored in subsequent papers.

 We expect about half of the ``simple'' sources to be star-forming galaxies, with the remaining half to be AGN. It is this simple sample for which we obtain multiwavelength data in this paper.The rest of the value-added processing described here is concerned only with this simple catalogue, and the value-added processing of the complex sources will be described in a future paper (Marvil et al., in preparation). }

In Table \ref{samplesize} we list the numbers of sources remaining at each stage of the value-added processing.

\begin{table*}
\begin{center}
\caption{Numbers of sources remaining after each stage of the value-added processing. }
\label{samplesize}
\begin{tabular}{lll}
\hline
Criterion	&	\# sources & Percentage of \\
& & simple source sample	\\
\hline
Initial source extraction & 220,102 & n/a\\
Restrict to simple sources & 178,921 & 100\\
CATWISE2020 Crossmatch* (10 arcsec) & 170,703 & 95\\
CATWISE2020 Crossmatch (3 arcsec) & 134,657 & 75\\
DES DR1 Crossmatch (2 arcsec) & 107,735 & 60\\
DES DR1 Crossmatch* (1 arcsec) & 91,811 & 51\\
Photometric redshift & 81,938 & 46\\
\hline
\end{tabular}
\medskip\\
Asterisked rows are shown for information but are not used in the subsequent selection step.
\rev{Photometric redshifts are taken from  
\citet{zou19},
\citet{zou19a},
and \citet{bilicki16}.}
\end{center}
\end{table*}

\subsection{Spectral Indices}
\label{spindex1}
Spectral indices of the simple sources are measured over the 288 MHz bandwidth of ASKAP using the Taylor term technique  described above. We  measure spectral indices  by calculating them from the Taylor terms at the peak pixel of each component, in the convolved data set. Note, as discussed above,  that this procedure differs from that in the \rev{initial} public data release, which we consider to be unreliable.

\rev{We also explored using the third Taylor term, which would measure spectral curvature, but found that very few sources had a measurable spectral curvature in the 288 MHz bandwidth of these observations. More importantly, we found that introducing a third Taylor term increased the uncertainty in the first two Taylor terms without increasing the accuracy, presumably because we are introducing a third free parameter which is primarily driven by noise. 
 }
 
\begin{figure}
\begin{center}
\includegraphics[width=8cm, angle=0]{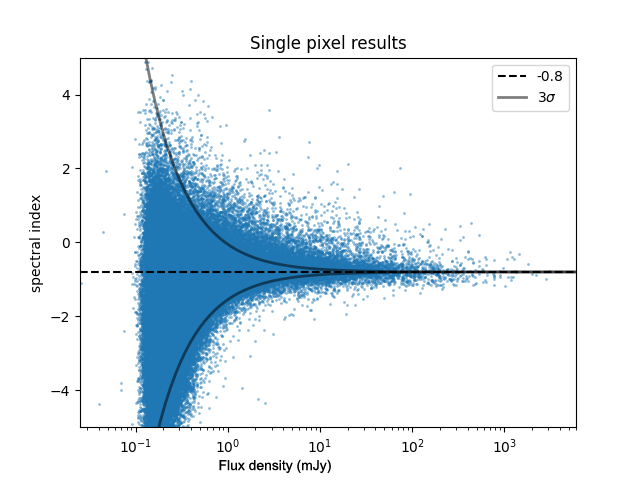}
\caption{The measured spectral index as a function of flux density. The two solid lines show the 3$\sigma$ uncertainty for a source of spectral index -0.8. Note the excess of sources with a positive spectral index, discussed in Section \ref{flatspectrum}.
 }
\label{spindex}
\end{center}
\end{figure}

The distribution of the resulting spectral indices as a function of flux density is shown in Figure \ref{spindex}. Based on the noise measured in the TT1 image, the $1\sigma$ spectral index uncertainty of a source with flux density $S$ mJy is $0.25 / S$. The spectral index of a 2.5 mJy source therefore has a standard error of $\sim$ 0.1, and spectral indices of  sources weaker than this will be increasingly uncertain.  

\begin{figure}
\begin{center}
\includegraphics[width=8cm, angle=0]{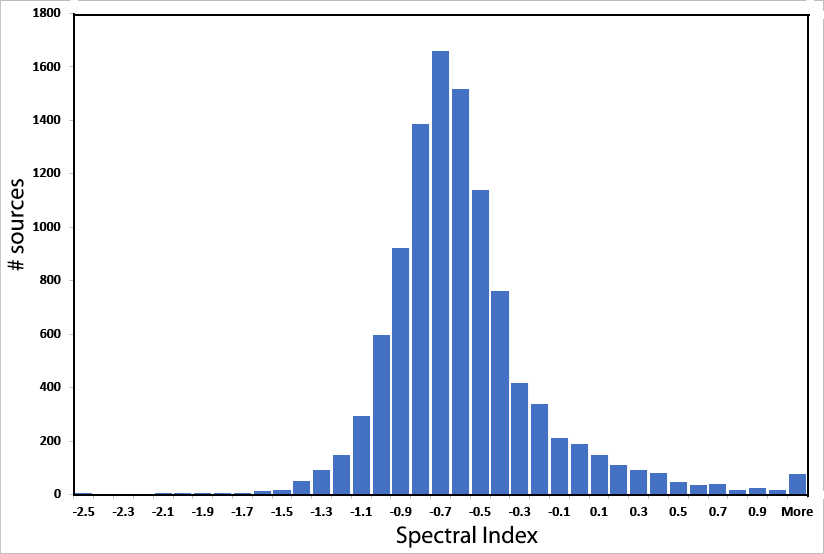}
\caption{A histogram of measured spectral index as a function of flux density, for the 10458 sources with  flux density $>$ 2.5 mJy.
 }
\label{spindexh}
\end{center}
\end{figure}

A histogram of the spectral indices for  the 10458 sources with  flux density $>$ 2.5 mJy is shown in Figure \ref{spindexh}. The peak is at a spectral index of --0.7, as expected for surveys of mJy radio sources, with a tail of steeper spectrum sources extending to $\alpha < -1.3$. Such ``ultra-steep spectrum sources'' are well-known in the literature \citep[e.g.][]{afonso11} and can be an indicator of high redshift sources. There is also an unexpected tail of sources with positive spectral indices. Such sources are also well-known \citep[e.g.][]{healey07} but are relatively rare. Here, however, they appear to constitute a significant fraction of EMU-PS sources. This is discussed further in Section \ref{flatspectrum}.

\subsection{Multi-wavelength Cross-identifications and Redshifts}

For cross-identifying simple sources, we use a simple nearest-neighbour cross-identification algorithm, and show below that this gives an acceptable completeness and false-ID rate. \citet{norris06} found that cross-matching with 3.6 $\mu$m {\it Spitzer} infrared data, and then cross-matching the infrared with optical, gives a lower false-ID rate than matching radio with optical directly. We therefore adopt this procedure here, and first match the radio against the W1 band (3.4 $\mu$m) of the  CATWISE2020 catalog \citep{marocco21}, hereafter referred to as ``CWISE'', and then cross-match the CWISE positions against the DES DR1 optical catalog \citep{DES}.

\begin{figure}
\begin{center}
\includegraphics[width=8cm, angle=0]{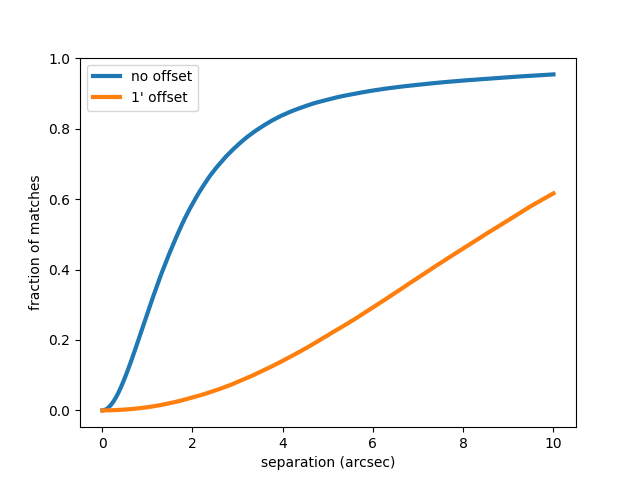}
\caption{The fraction of simple radio sources (as listed in Table \ref{samplesize}) matched with a CWISE source as a function of separation, both for unshifted data and for data shifted by one arcmin.
 }
\label{xmatch}
\end{center}
\end{figure}

We measured the number of cross-matches  between the radio and the infrared as a function of separation, and then estimated the false-ID rate by shifting the radio positions by 1 arcmin and then repeating the cross-match. The result is shown  in Figure \ref{xmatch}. The choice of an optimum search radius depends on the application (i.e., whether the goal depends on maximising the number of cross-matches or minimising the number of false-IDs). In producing the cross-matched catalogue we include all cross-matches up to a search radius of 10 arcsec so that users can choose their optimum search radius, but for further work herein we limit our analysis to a maximum search radius of 3 arcsec, at which we find an 8\% false-ID rate and a 75\% total cross-match rate (which includes the false-IDs). The resulting numbers of sources are listed in Table \ref{samplesize}.

Because of the high numbers of faint CWISE sources, we also explored the effect of introducing a cutoff in the CWISE flux densities, so only the brighter sources would be cross-matched to radio sources, but found that had a negligible effect on the false-ID rate, while significantly reducing the number of true IDs, and so no cutoff is used.
 
\begin{figure}
\begin{center}
\includegraphics[width=8cm, angle=0]{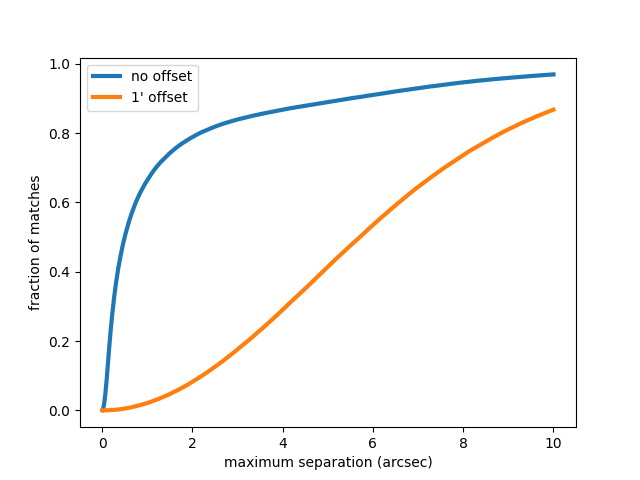}
\caption{The fraction of radio sources with a CWISE position matched with a DES DR1 source as a function of separation, both for unshifted data and for data shifted by one arcmin.
 }
\label{xmatchDES}
\end{center}
\end{figure}

To cross-match the CWISE IR positions against DES optical positions, we again explored the false-ID rate and the total-ID
rate as a function of search radius, and show the results in Figure \ref{xmatchDES}. As a result, we
adopt an  search radius of 2 arcsec. 
The resulting numbers of sources are listed in Table \ref{samplesize}. 

There is no major spectroscopic redshift survey covering the EMU-PS field, but  a large number of photometric redshifts are available from \cite{bilicki16} (using their ``main'' catalogue), and from \citet{zou19, zou19a}, and  we also include those in the catalogue. Throughout the rest of this paper, redshifts given without a citation refer to these redshifts used in the EMU-PS catalogue.

\subsection{Astrometric Precision}
\begin{figure}
\begin{center}
\includegraphics[width=8cm, angle=0]{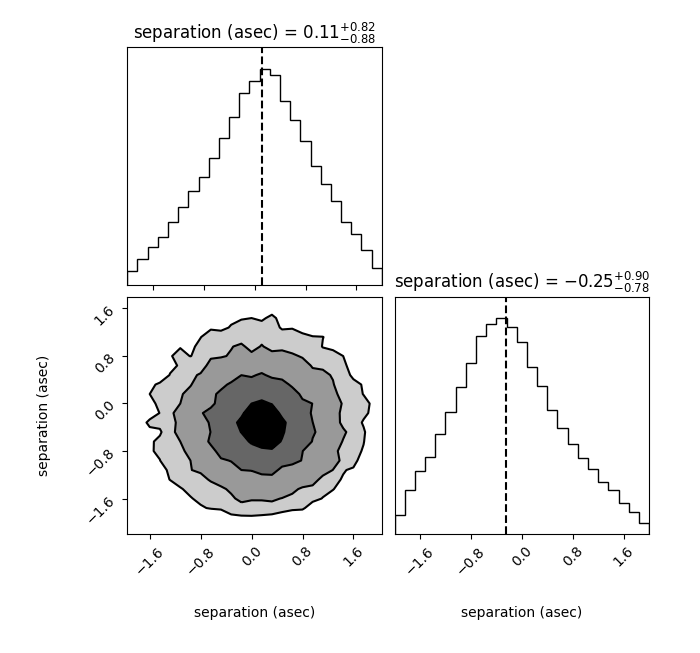}
\caption{A plot showing the difference in position of radio sources compared to the matching CWISE source in the W1 band, showing a mean offset of $\sim$ 0.3 arcsec, which is small compared to the 18 arcsec resolution  of the convolved data. The horizontal axis is Right Ascension and the vertical axis is Declination.
 }
\label{astrometry}
\end{center}
\end{figure}

For each source that was cross-matched with a CWISE catalogue source,  we measured the offset in position, as a check  on the  precision of the positions of the radio components. The result is shown in Figure \ref{astrometry}, showing a mean offset of $\sim$ 0.3 arcsec, which is small compared to the 18 arcsec resolution of the convolved data. The positions in the catalogue have not been corrected for this insignificant offset.

\rev{
\subsection{Flux Density Accuracy}
To estimate the flux density accuracy, we select EMU-PS sources stronger than 6 mJy (the minimum flux density for sources in the SUMSS \citep{sumss} catalogue), and cross-match them to SUMSS sources using a 3 arcsec search radius, which selects about 50\% of the SUMSS sources, and tends to exclude the very extended SUMSS sources. We then calculate the ratio of peak fluxes in the EMU-PS and SUMMS catalogs. The result is shown in Figure \ref{fluxratio}.

Ideally we would convolve the EMU-PS to the 45 arcsec resolution of SUMSS and then repeat the source extraction, but then it would not be matched to the EMU-PS value-added catalogue. Because we have not done this convolution, some SUMSS peak flux densities are boosted by components which are included in the SUMSS beam but not in the EMU-PS beam. This  increases the scatter of the ratios, so that the measured scatter in the ratio is an overestimate of the uncertainty in the EMU-PS flux density scale.

We note the following features of Figure \ref{fluxratio}.
\begin{itemize}
\item The EMU-PS central frequency of 944 MHz differs from the SUMSS central frequency of 843 MHz, and, assuming a spectral index of -0.8, we expect the peak of the distribution to occur at a ratio of $\sim  (944/843)^{0.8} = 1.09 $ as observed.

\item The histogram is more extended on the right, presumably because of the larger size of the SUMSS beam which will boost the SUMSS peak flux as discussed above.
\item The  left of the histogram is approximately Gaussian with a standard deviation $\sim$ 0.12. 
\end{itemize}

We therefore estimate our flux density scale uncertainty  for strong sources to have a maximum value  of $\sigma \sim$ 12\%. To this should be added in quadrature the estimated flux density scale standard error for SUMSS of 3\% \citep{sumss}.

 The measured flux density of weaker sources will be degraded by a factor of 1/SNR, where SNR is the local signal-to-noise ratio. As we have rejected sources from the EMU-PS catalog with SNR$<$5, this may add an uncertainty of up to 20\% (to be added in quadrature) to the quoted flux densities of weak sources.

\begin{figure}
\begin{center}
\includegraphics[width=8cm, angle=0]{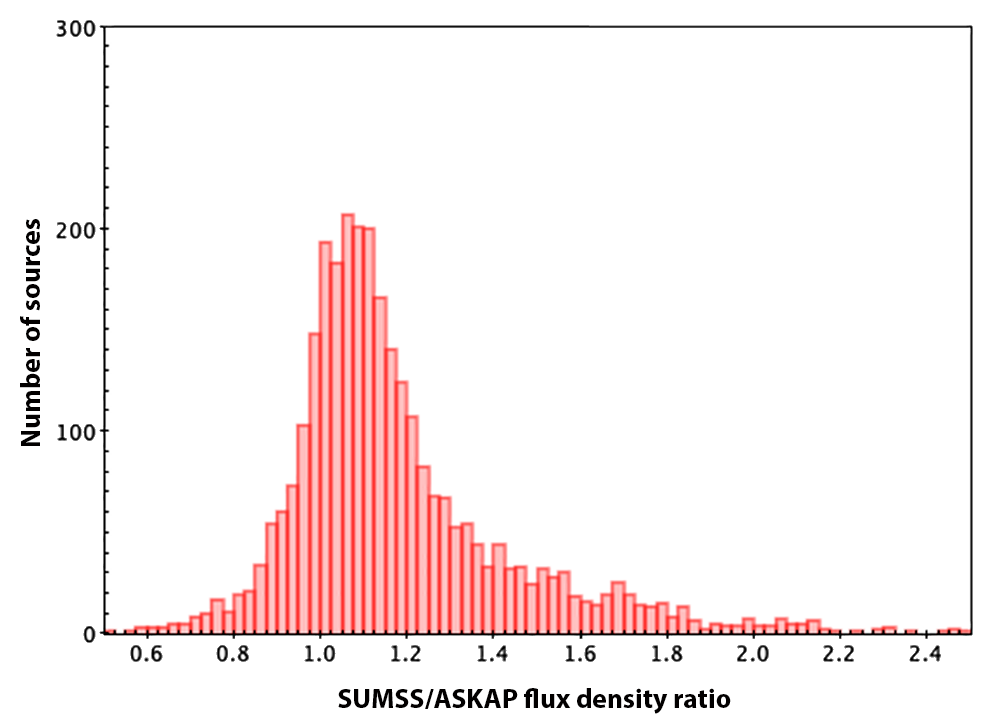}
\caption{The ratio of peak flux densities between EMU-PS and SUMSS for simple sources with EMU-PS flux densities $>$ 6 mJy, and with catalogued positions within 3 arcsec.
 }
\label{fluxratio}
\end{center}
\end{figure}
}

\section{Results}

\subsection{Data Summary and Access}
\label{public}
\label{proprietary}

The EMU Pilot Survey has produced an image of about 270  \sqdeg\ of the radio sky at 944 MHz, with a spatial resolution of $\sim$ 11--13 arcsec and an rms sensitivity of $\sim$25--30 \ujybm. 

\begin{figure*}
    \centering
    \includegraphics[width=0.9\textwidth]{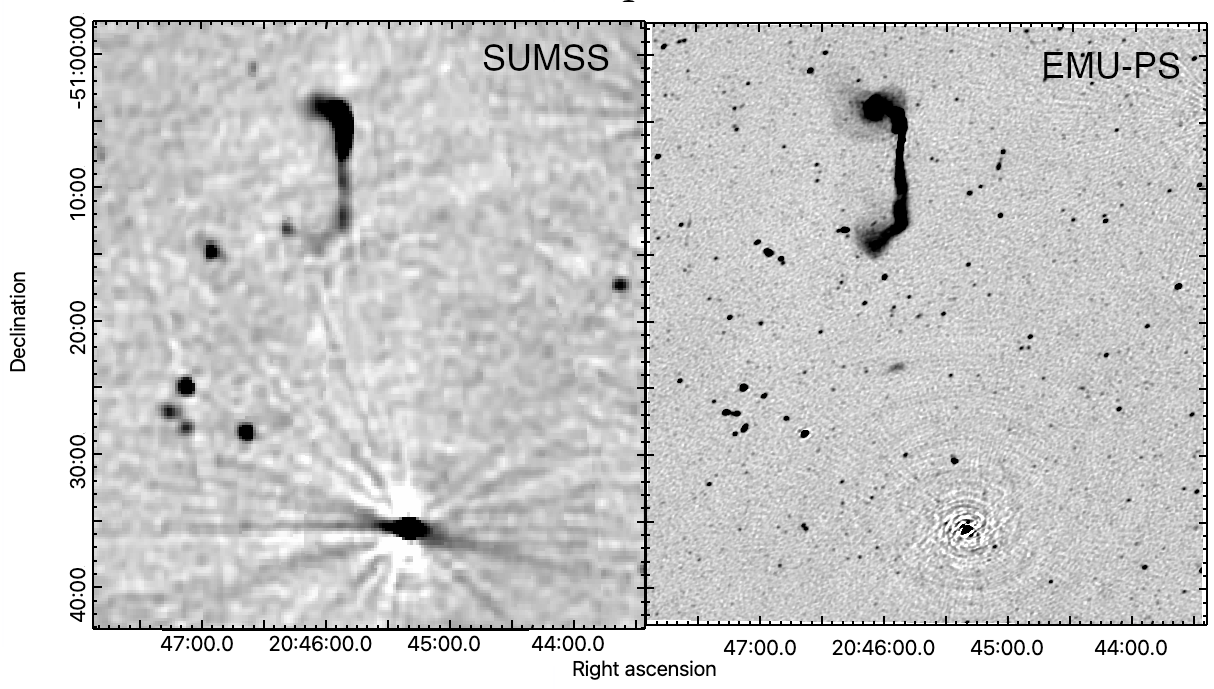}
    \caption{A typical section of the survey field, covering about 0.3 \sqdeg (or about one thousandth of the area of the EMU Pilot Survey) which contains about 250 radio sources). On the left is the SUMSS image \citep{sumss} and on the right is the EMU-PS image. Prominent in this image is the Giant Radio Galaxy ESO 234-68. The maximum flux density of ESO 234-68 in the EMU-PS image is 58.8 mJy/beam, and that of the strong source at the bottom of the image (PMN J2045-5135) is 1.06 Jy/beam. The rms of the EMU-PS image is 25-30 \ujybm, and that of the SUMSS image is $\sim$1.25 mJy/beam.}
    \label{typical}
\end{figure*}

A problem with large surveys is that it is difficult to convey the scale and depth of the image in a journal paper. Figure \ref{merged} shows the entire native resolution image, and  Figure \ref{typical} shows a random section of it, which covers about one thousandth of  the area of the EMU Pilot Survey. An interactive interface to the image of the entire survey field in HiPS format  is available on \url{http://emu-survey.org}.  

After observing, \rev{processing, and validation by the EMU survey team,} the data from each observation are placed on the CSIRO ASKAP Science Data Archive (CASDA) data server, and made available 
to the public as described below. 
These data consist of all the data from each day's observations, known as a ``tile'', including images and tables of extracted components and islands. We call this catalogue the EMU Pilot \rev{Initial} Public Data Release. The validation metrics and flags are associated with each tile and are fully queryable via Table Access Protocol (TAP).

 The data are \rev{then} processed by merging tiles into a common image covering the whole field of the EMU-PS.  The resulting data release of this image is called the ``native'' value-added data release.
 
\rev{As described in Section \ref{value-added},
we then
smooth the native resolution image to a constant resolution of 18 arcsec, perform source extraction, and separate the resulting catalogue of 220,102 components into simple, single, components (81\%), and more complex sources (19\%). We also perform cross-identifications of the simple sources with other available multiwavelength products.

 }
 
\begin{figure}
\begin{center}
\includegraphics[width=8cm, angle=0]{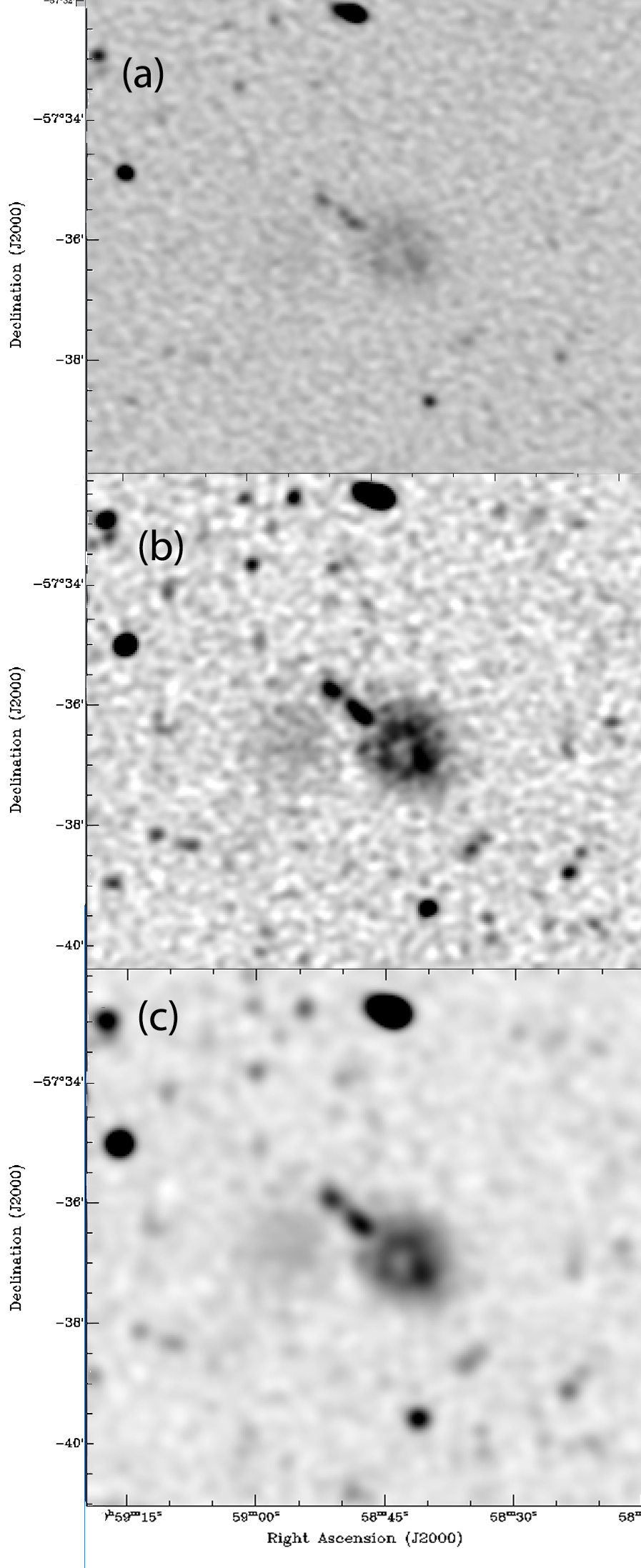}
\caption{A sample of the final image, showing the three  data products on a region, covered by three tiles, containing two of the ``Odd Radio Circles''\citep{orc}: (a) the \rev{initial} public data release from a single tile (SB9351) (resolution 11$\times$13 arcsec, rms = 40 \ujybm, (b) the added-value ``native'' data release with 11$\times$13 arcsec resolution, from the merged tiles, rms = 25 \ujybm, and (c) the added-value ``convolved'' data release with 18 arcsec resolution, rms = 25 \ujybm. The peak flux density in this image is 4.6 mJy/beam.
}
\label{sample}
\end{center}
\end{figure}

We call this science-ready data set the ``Convolved'' data set.  
The resulting
image for an area of sky covering an object of interest is shown in Figure \ref{sample}, which shows the three  data products: 
the \rev{initial} public data release, the added-value ``native'' data release with 11--13 arcsec resolution, and the added-value ``convolved'' data release with 18 arcsec resolution.

\rev{ All three  data products from the EMU-PS (the \rev{initial} public data release, the added-value ``native'' data release with 11--13 arcsec resolution, and the added-value ``convolved'' data release with 18 arcsec resolution)
 are released via the CASDA data server described below. The \rev{initial} public data release is currently available in the public domain, but the two added-value data releases are available only to EMU members for a proprietary period of one year from the date of publication of this paper, after which they will be released into the public domain. However, EMU is an open collaboration, and other astronomers are welcome to join the project, and access the proprietary data, provided they agree to the EMU data and publication policies. 
}

\rev{The EMU-PS initial public release data in the CSIRO ASKAP Science Data Archive (CASDA) are open to the public domain. To download data from CASDA, users need to obtain a CASS Online Proposal Applications and Links (OPAL) account\footnote{ \url{https://opal.atnf.csiro.au/}}.
}

CASDA is described in detail by \cite{Chapman2017} and \cite{Huynh2020}. In brief, CASDA is implemented across two data centres, the Pawsey Supercomputing Centre in Perth and the CSIRO data centre in Canberra. So-called ``backend" functions such as deposit, storage and data access are implemented at Pawsey, while the ``frontend'' functions such as the user-interface and authentication are implemented at the CSIRO data center. 

The simplest way to access the data is via the CASDA web user-interface. From the CASDA webpage\footnote{\url{https://casda.csiro.au}}, select ``Access CASDA via the Data Access Portal'', to be taken to the Observation Search user-interface. EMU-PS data can be obtained by searching for ``Released'' data under project code AS101. 
\rev{
EMU-PS data have Digital Object Identifiers (DOIs) which provide a persistent resolvable link to the data.  Table \ref{DOI} gives the DOI for each data product discussed in this paper. The DOI links to the collection page; from there click on ``files" and select the files to download. 

\begin{table*}
\begin{center}

\caption{Available data products, including Digital Object Identifiers (DOIs) that can be used to access the data described in this paper. }
\label{DOI}
\begin{tabular}{ll}
\hline
Description	&	DOI\\
\hline
Initial Public Data Release images & \url{https://doi.org/10.25919/5f27829da27ba}\\
Initial Public Data Release  catalogues & \url{https://doi.org/10.25919/5f27823fa37e2}\\
Added-value Data Release native resolution image & \url{https://doi.org/10.25919/exq5-t894}\\
Added-value Data Release convolved image (TT0) & \url{https://doi.org/10.25919/exq5-t894}\\
Added-value Data Release convolved image (TT1) & \url{https://doi.org/10.25919/exq5-t894}\\
Added-value Data Release convolved catalogue & \url{https://doi.org/10.25919/exq5-t894}\\
\hline
\end{tabular}
\end{center}
Notes:
\begin{itemize}
    \item The initial public data release is immediately available, but the value added releases are available only to members of the EMU collaboration for one year from the data of publication of this paper, after which they become public. 
\item All catalogues contain island and component information, and, for the added-value catalogue, cross-identifications and redshifts, where available. 
\item TT0 and TT1 refer to Taylor Term 0 image (total power) and Taylor Term 1 image (TT0 $\times$ spectral index
\end{itemize}

\medskip
\end{table*}
}

CASDA also implements several Virtual Observatory services to maximise the usability and interoperability of ASKAP data products, and allow for automated scripted access. For example, the table access protocol (TAP) can be used to search for EMU-PS observations under project code AS101, using an application such as TOPCAT \citep{taylor05} or Aladin \citep{aladin, aladin1}. A CASDA module has recently been added to the Python astropy astroquery\footnote{\url{https://astroquery.readthedocs.io}} package. Using this Python API, the EMU-PS images can be accessed and downloaded with a cone search of the EMU-PS pointings.

All public data (tables and images, and $u,v$ data) are available from CASDA\footnote{ \url{http://hdl.handle.net/102.100.100/164555?index=1}}
and a listing of all ASKAP observations is on the \rev{Observation Management Portal (OMP)\footnote{ \url{https://apps.atnf.csiro.au/OMP/index.jsp}}}. {\bf OMP allows the user to select observations by several parameters including date, SBID (listed in Table \ref{observations}), or project name (AS101 for EMU).}


\subsection{Sensitivity \rev{to Compact Sources}}
The EMU-PS survey reaches a typical sensitivity of 25-30 \ujybm\ rms. This is about a factor of two above the calculated thermal noise sensitivity ($\sim$ 13 \ujybm), which we tentatively attribute to the following causes.
\begin{itemize}
\item Timing errors in the correlator  cause a significant fraction of data (30-50\%) to be flagged, resulting in a loss of data. Work is in progress to identify and eliminate the cause of this problem.
\item The data calibration processes are in a preliminary state. By the time of the final EMU survey, we expect to have developed a sky model 
which will be used to calibrate the data and remove strong sources prior to cleaning.
\item A dynamic range problem, which is currently being addressed, causes diffraction patterns around strong sources.
\item The primary beam correction  assumes a Gaussian profile across each PAF beam. This is being replaced by a profile based on holographic measurements which will be beam-specific.
\item The lack of direction-dependent calibration, which we hope to address in the future.
\end{itemize} 

After correcting these errors, and including a confusion noise of about 9 \ujybm, we expect the full EMU survey (conducted at a centre frequency of $\sim$ 944 MHz) can potentially reach an rms of about 17.5 \ujybm. 
\rev{
\subsection{Sensitivity to Extended Emission}
As well as its high sensitivity to compact sources, the survey also has high sensitivity to extended low surface brightness emission, because of the large number of short spacings in the ASKAP array.

In Figure \ref{scaled-sensitivity} we show a plot of the sensitivity of EMU-PS as a function of spatial scale, obtained by running simulated observations with different tapers, producing different beamsizes.

\begin{figure*}
\includegraphics[width=17cm]{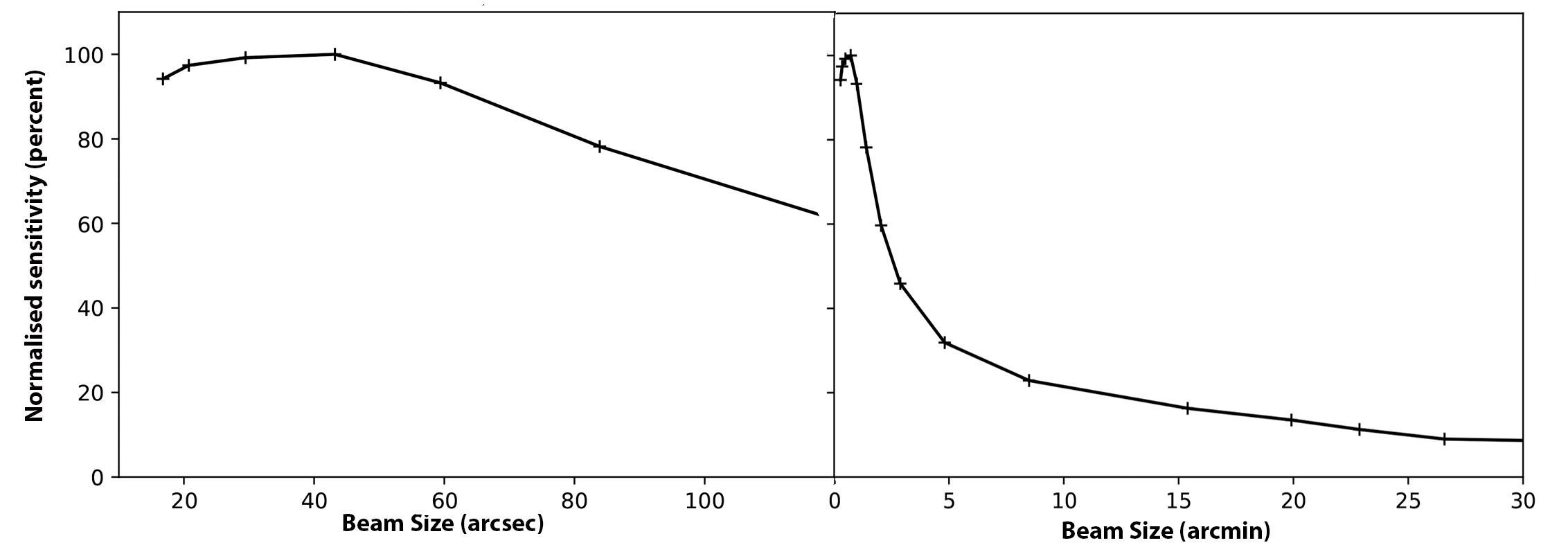}
\caption{\rev{The sensitivity of EMU-PS as a function of spatial scale.  The plot was made using visibility data from a single beam and pointing of an interleaved observation (~2 hour observation, 288 MHz bandwidth, scaled to the EMU-PS observing frequency of 944 MHz) which was filled with Gaussian noise and various uv-tapers were applied to shape the beam size. We then measured the image noise (effectively the sensitivity at the scale associated with the uv-taper). The two plots show the same result over different ranges of spatial scale.
}
}
\label{scaled-sensitivity}
\end{figure*}

The sensitivity of 25-30 \ujybm\ at the native resolution of 11-13 arcsec is almost unchanged at the convolved resolution of 18 arcsec, and continues at a similar level beyond the  45 arcsec resolution of SUMSS \citep{sumss}, which has a median rms sensitivity of 1.27 mJy/beam. The effect of this high sensitivity to low surface brightness emission is demonstrated in Section \ref{science}.
}

\subsection{Source Counts and Confusion}

\begin{figure*}
\includegraphics[width=18cm,clip]{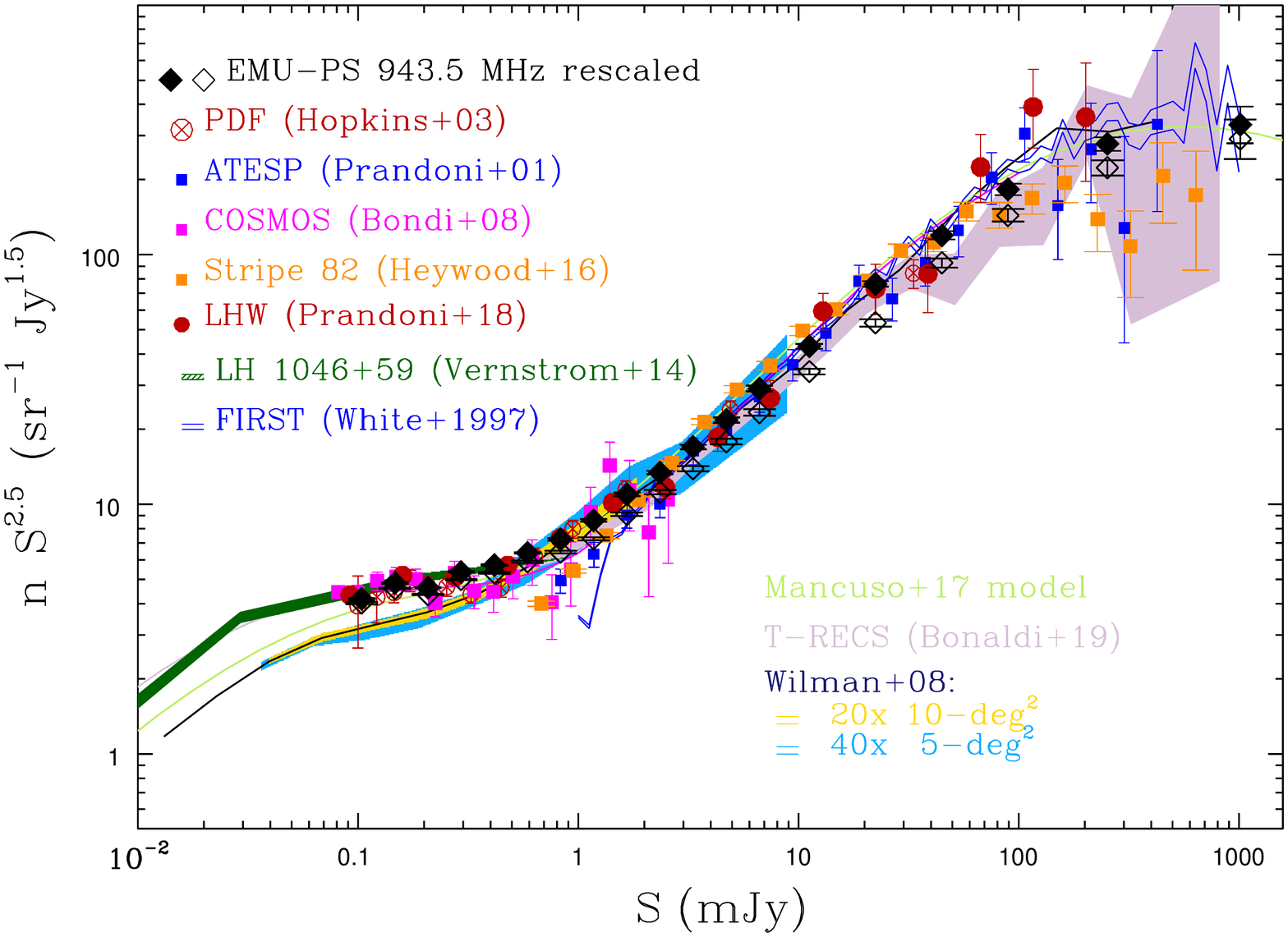}
\hfill
\caption{Normalized differential source counts derived from the 270 deg$^2$ EMU-PS survey for the {\it island} catalogue (black filled diamonds) and for simple sources only (black empty diamonds). The counts have been rescaled from 943.5 MHz to 1.4 GHz by assuming $\alpha$=-0.7. Also shown for comparison are the counts derived from 1.4 GHz $>$ degree-scale surveys (symbols and colors as  indicated in the figure). Vertical bars represent Poissonian errors on the normalized counts. \rev{Systematic errors due to incompleteness corrections and spectral index assumptions are approximately included in the size of the plotted symbols.} The result of the $P(D)$ analysis performed by \citet[][rescaled from 3 to 1.4 GHz by assuming $\alpha$=-0.7]{vernstrom14} is indicated in dark green. The black solid line represents the predicted counts from 200 sq. degr. of the S3-SEX simulations  (\citealt{wilman08}). The \rev{light  blue and yellow} shaded areas illustrate the predicted cosmic variance effects for survey coverages of 5 and 10 sq. degr. respectively (obtained by splitting the  S3-SEX simulation in forty 5-deg$^2$ and twenty 10-deg$^2$ fields respectively). The 25 deg$^2$ medium tier of the more recent T-RECS simulations (\citealt{bonaldi19}) is represented by the purple shaded area. Finally, the \citet{mancuso17} radio source evolutionary model is shown by the light green line. }
    \label{fig:sourcecounts}
\end{figure*}

 Figure~\ref{fig:sourcecounts} shows the differential source counts normalized to a non evolving Euclidean model (n $\propto S^{2.5}$) obtained from the EMU-PS catalogue (black symbols), rescaled from 943.5 MHz to 1.4 GHz by assuming $\alpha$=-0.7 . Two counts' determinations are shown: one referring to the {\it island} catalogue, where components of complex sources are merged together (filled diamonds) and one referring to simple sources only (empty diamonds). The source counts are corrected for both Eddington bias \citep{eddington13,eddington40} and resolution bias (i.e. the  incompleteness introduced by the fact that a larger source of a given total flux density will drop below the signal-to-noise threshold of a survey more easily than a smaller source of the same total flux density). This is done following standard recipes in the literature (see e.g. \citealt{prandoni01,prandoni18,mandal21}). 
 
 Figure~\ref{fig:sourcecounts} shows for comparison some of the widest-area samples available to date at 1.4 GHz. This includes sub-mJy surveys covering $>$1 deg$^2$ regions, like PDF \citep{hopkins03},  VLA-COSMOS \citep{bondi08} and the 6 deg$^2$ \rev{Westerbork} mosaic covering the Lockman Hole region \rev{\citep[LHW:][]{prandoni18}}, as well as shallower ($> 1$ mJy) but larger ($\gg$10 sq. degr.) surveys like  ATESP \citep{prandoni01}, SDSS Stripe 82 \citep{heywood16} and FIRST \citep{white97}. 
Also shown are simulated source counts derived by combining evolutionary models of either classical radio loud (RL) AGN or radio source populations dominating the sub-mJy radio sky, namely star forming galaxies (SFG) and low-luminosity  AGN (LLAGN). In particular we show the 1.4 GHz counts derived from the recent modeling of \citet[light green solid line]{mancuso17},  the T-RECS 25 deg$^2$ medium tier simulation (\citealt{bonaldi19}), as well as different realizations obtained from the S3-SEX simulated catalogue (\citealt{wilman08}): $1\times 200$ deg$^2$  (black solid line), $20\times 10$ deg$^2$ regions (\rev{yellow} shaded area) and $40\times 5$ deg$^2$ regions (light blue shaded area).  

Figure~\ref{fig:sourcecounts} clearly shows that the source counts derived from the EMU-PS {\it island} catalogue nicely match previous counts and are in good agreement with the most recent models/simulations (\citealt{mancuso17,bonaldi19}). Even more interestingly they provide very robust statistics all the way from $\sim 0.1$ mJy to $>1$ Jy, something which could only be achieved in the past by combining deeper (but smaller) surveys with larger (but shallower) surveys. Finally it is interesting to note that the counts derived from simple sources only (black empty diamonds) fall well below the full counts (black filled diamonds) at bright fluxes. This is not surprising  as we expect a large contribution from multi-component RL AGN at flux densities $\gg 1$ mJy. On the other hand no significant difference is observed at sub-mJy fluxes, confirming that this flux regime is dominated by SFG and LLAGN.

\nocite{prandoni18,hopkins03, prandoni01, bondi08, heywood16, white97, vernstrom14, mancuso17,bonaldi19}

\subsubsection{Source Confusion}

We estimate the source confusion noise, and instrumental noise using the probability of deflection, or $P(D)$ technique \citep{Scheuer57}. The $P(D)$ distribution of an image is the distribution of pixel intensities (Jy beam$^{-1}$) which depends on the underlying source count, shape of the beam, and the instrumental noise \citep[see][for a detailed description of the method]{vernstrom14}. 

\begin{figure}
\includegraphics[width=\columnwidth]{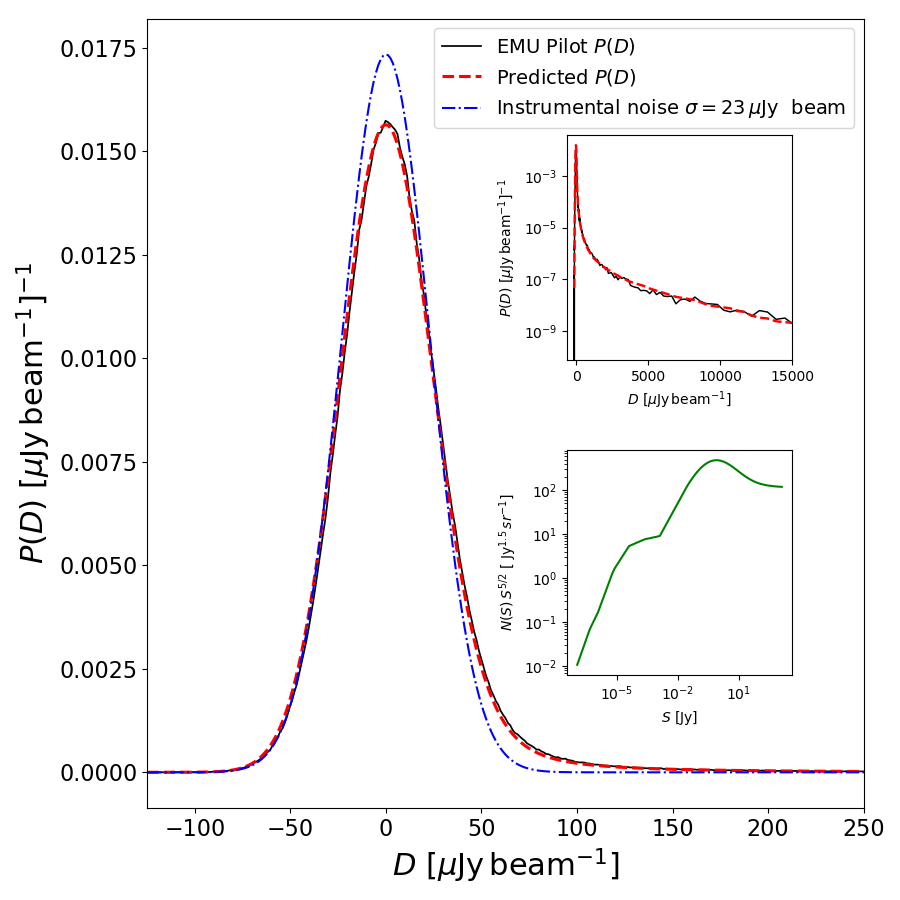}
\caption{The EMU-PS preliminary $P(D)$ distributions. The solid black line is the probability distribution made from sections of the pilot away from bright sources.  The upper right inset shows bright flux density tail of the $P(D)$ distributions.  The blue dot-dashed line shows a Gaussian noise distribution of $\sigma= 23 \mu$Jy beam$^{-1}$. The red dashed line shows the predicted or model $P(D)$ generated from the source count shown in the lower right inset.}
\label{pofd}
\end{figure}

The $P(D)$ method assumes a Gaussian distribution for the instrumental noise and can therefore be affected by imaging artefacts, such as those found around bright sources. We computed the histogram of pixel intensities for the pilot survey image  by selecting regions of pixels devoid of any image artefacts, as well as any complex diffuse or extended emission. Rather than a full source count fitting analysis, which is beyond the scope of this paper, we take the deep $P(D)$ source counts derived in \citet{vernstrom14} and scale it to a frequency of $944\,$MHz using $\alpha =-0.7$. We take the average beam sizes from the individual beams and find $B_{\rm maj}=12\,$ arcsec and $B_{\rm min}=10\,$ arcsec, while using an image of the ``dirty'' synthesized beam for sources below the clean limit (approximated at $S_{\rm clean}=200\, \mu$Jy beam$^{-1}$). We use an average instrumental noise value of $\sigma=23\, \mu$Jy beam$^{-1}$. The image $P(D)$, noise distribution, and model $P(D)$ can be seen in Fig.~\ref{pofd}. 

Without any additional fitting or changes to the parameters or source count, we find very good agreement between the image and model $P(D)$ distributions. The noise-free model $P(D)$ provides an estimate of the confusion noise in the field of $\sigma_{\rm conf}=5\, \mu$Jy beam$^{-1}$. This quick test shows through independent means that the instrumental noise estimate of $20$ to $25\, \mu$Jy beam$^{-1}$ is accurate. Furthermore, the fact that the scaled source count model provides a good match to the image is a confirmation of accurate source flux densities in the pilot data, and that confusion noise is not a significant factor for our scientific investigations, even when considering the dirty beam sidelobe confusion noise. At the same time, the observations of P(D) are sensitive enough to probe far below the source populations that we can directly detect.

\section{Preliminary Science Results}
\label{science}

\subsection{Peculiar Radio sources}

Many unusual radio sources are found in the EMU Pilot Survey.

\begin{figure*} 
    \centering
   \includegraphics[width=8cm]{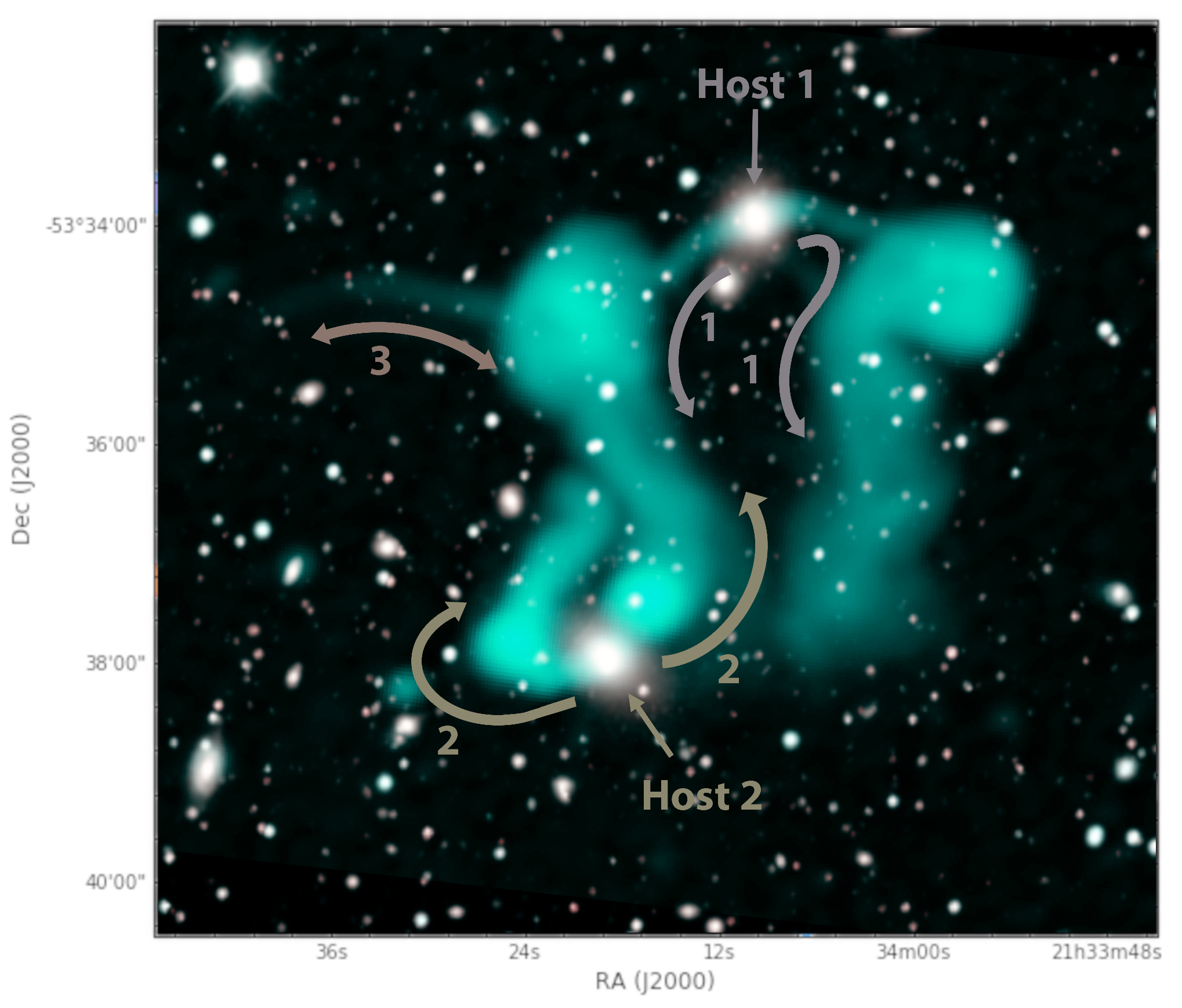}
    \includegraphics[width=8.8cm]{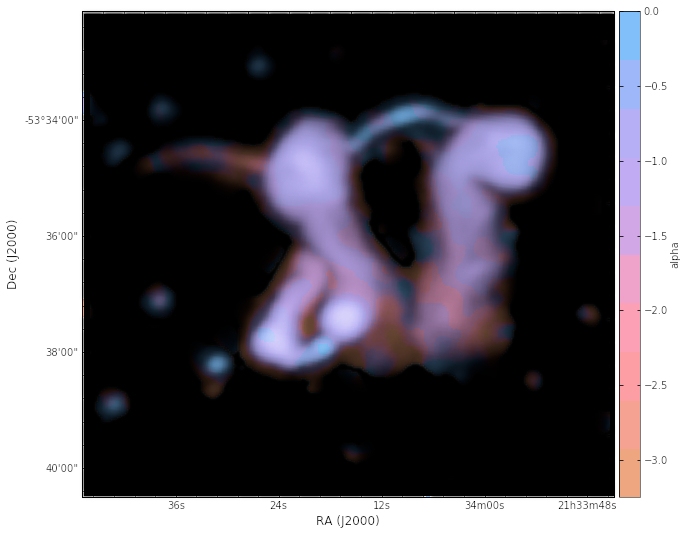}
    \caption{A peculiar radio source found in the EMU Pilot Survey, consisting of a group of distorted radio components, collectively known as PKS 2130--538, and nicknamed ``the dancing ghosts''. The two host galaxies ($z \sim 0.077$) are seen at the centre of the narrow jets \rev{(shown with numbers in the Figure to indicate their putative host)} which expand into diffuse lobes, probably bent by interactions. 
    On the left is the total intensity greyscale image (shown in turquoise), superimposed on a background of the DES optical image, assembled from the r, g, and i images. 
    On the right is the total intensity image of PKS~2130-538, colour-coded by spectral index.  The unconventional colour scheme was constructed using sequential colours on the ``colour-wheel"  \citep[e.g.][]{itten70}. 
    The colours were fixed in luminosity, i.e.,  fixed to be constant in luminosity-chroma-hue color space \citep{Ferrand19}. In this way, the brightness level on the image represents only the total intensity values. The colour bar indicates the spectral index at a single fixed intensity. Since the spectral index map in this color scheme was multiplied by the total intensity map, darker versions of colours are associated with fainter regions in the data. The peak flux density in this image is 103 mJy/beam.}
    \label{fig:ghosts} 
\end{figure*}

The source PKS 2130-538, shown in Figure \ref{fig:ghosts} has been previously identified as a complex source \citep[e.g.][]{ekers70,schilizzi75,jones92}, and as two radio galaxies (G4Jy 1704 and G4Jy 1705) in the G4Jy Sample \citep{white20a,white20b}. However, no previous image shows the wealth of detail and low surface brightness emission  seen in Figure \ref{fig:ghosts}.
It consists of the radio lobes of two host galaxies, one of which (\rev{``Host 1'':} 2MASX J21341775-5338101)  is the bright galaxy at the centre of the curved northern radio bridge, at a redshift of 0.0781. This  is the brightest  galaxy of the cluster Abell 3785.
The other host galaxy (\rev{``Host 2'':} 2MASX J21340666-5334186) is the bright galaxy near the southeast  end, at a redshift of 0.0763.  

The spectral index image helps to isolate the contributions from these two hosts. In the north, there is a very flat spectrum region ($\alpha \sim 0$) at the position of \rev{Host 1}, connecting to relatively flat $\alpha \sim -0.4 - -0.5$ jets \rev(``1''). These then connect to the large bright regions of $\alpha \sim -0.6 - -0.7$, steepening sharply down the tails to the south to at least -1.5, beyond which the spectra become more uncertain.  All of this is consistent with the behavior of bent-tail galaxies.  In the eastern half, there is a dramatic change in spectral index at the position of the emission associated with the second host; it has  its own flat core and steeper lobe/tail structures.   Although the overall emission comes from two distinct hosts, it is unclear whether there is an interaction between them, or merely a superposition.  An additional curiosity is the thin stream of emission \rev{``3''} extending eastward from the NE bright region;  it has a median spectral index of -2.1, and both its dynamical origins and particle history don't fit naturally into existing radio galaxy models.

\begin{figure} 
    \centering
    \includegraphics[width=8cm]{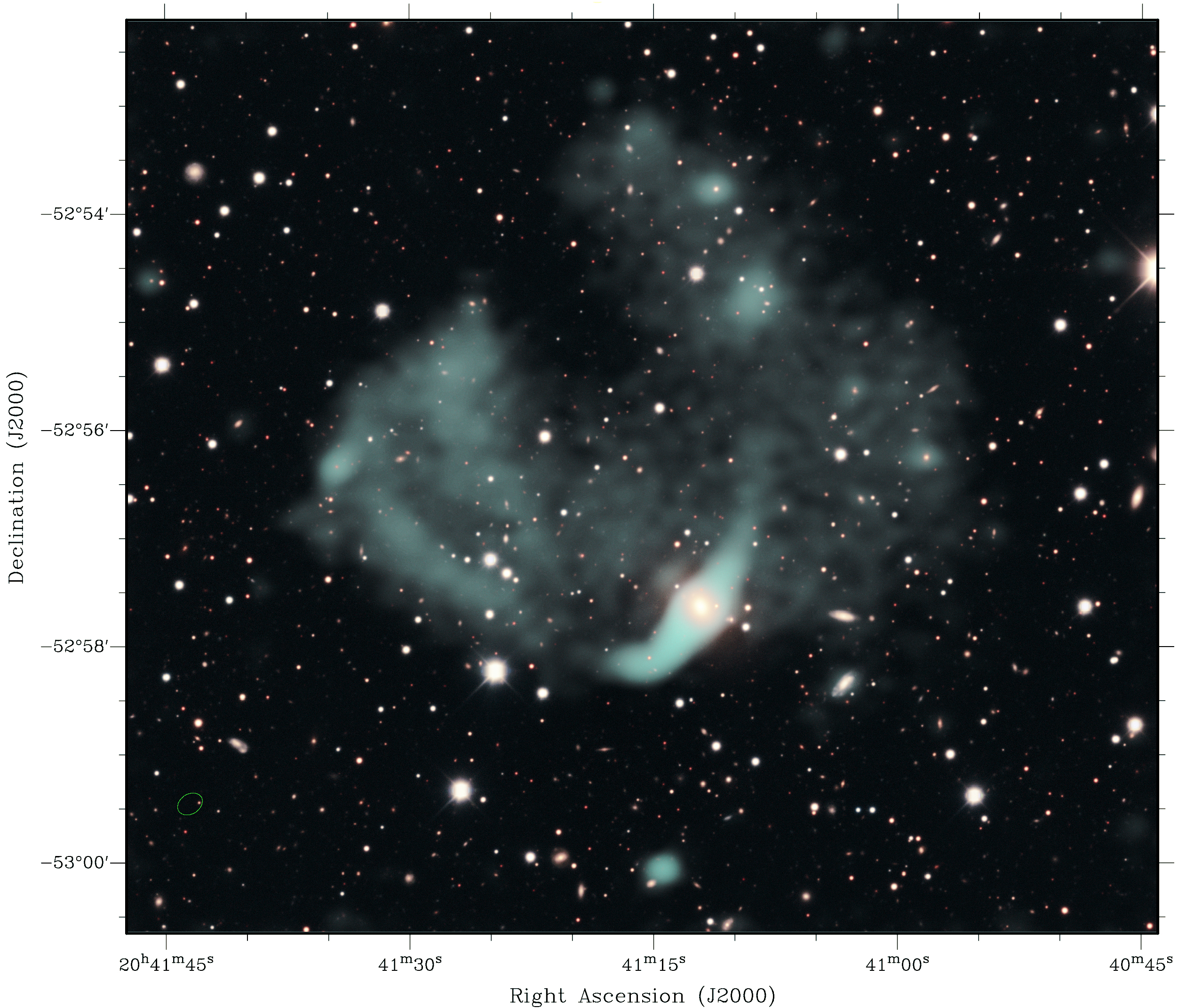}
    \caption{Another peculiar radio source found in the EMU Pilot Survey: a double lobed radio AGN, known as PMN J2041--5256, with a curious ``double'' bent tail. The radio data from EMU-PS has been ``stretched'' to show the faint emission, and then coloured turquoise, and adjusted to emphasise the double tail. DES g-, r-, and i-band data are combined to form the background, which is combined with the radio data using a layer mask in GIMP. Embedded in the tails are several radio sources that may be unrelated to the tailed galaxy. The peak flux density in this image is 58.3 mJy/beam.}
    \label{fig:doubletail} 
\end{figure}

The source PMN J2041-5256, shown in Figure \ref{fig:doubletail}, is a double lobed radio AGN, associated with the host galaxy WISEA J204112.05-525737.7 at a redshift of z=0.048. Previous radio data \citep{sumss,gregory94} show only an indistinct extended source corresponding to the nucleus. Its jets \rev{(shown with numbers in the Figure to indicate their putative host)} are presumably being bent by intra-cluster winds, but the morphology is much more complex than normal bent-tail galaxies. The eastern jet \rev{(``2'')}  is bifurcated, whilst the western jet \rev{(``1'')} breaks down into a number of blobs, accompanied by a large diffuse area of emission to the west of the source. One possibility is that its relative motion with respect to the intracluster medium has a large component along the line of sight;  the bifurcated tail and western diffuse extensions would then be more typical of structures seen in bent-tail galaxies, but seen here in projection. If the direction of motion were ~$\sim$20 degrees from the line of sight, the entire source length would be $\sim$750~kpc, among the larger bent-tail sources.  

\begin{figure}
\begin{center}
\includegraphics[width=8cm, angle=0]{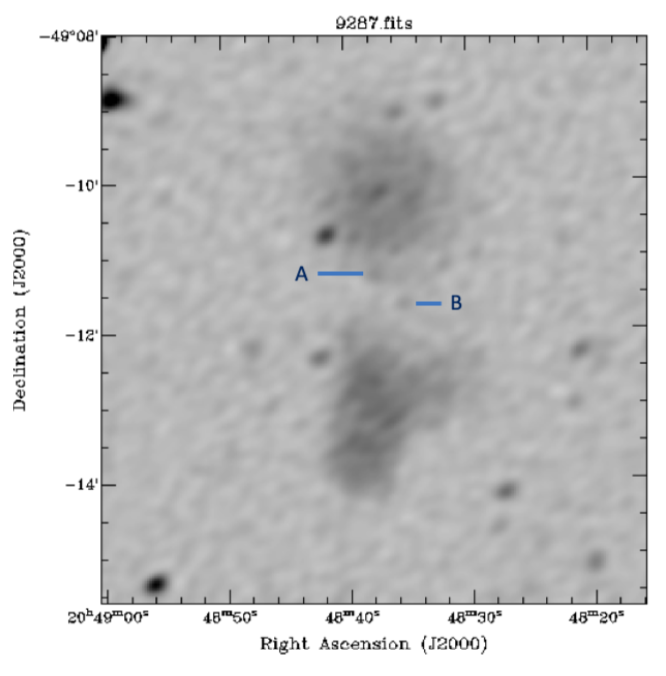}
\caption{The ``Smoking Gun'' Galaxy EMU PS J204835.0--491137 consists of the two diffuse radio clouds seen in this image. These are presumably the remnants of a classical double-lobed radio galaxy in which the central engine has switched off. The labels A and B indicate two possible host galaxies, discussed in the text. The peak flux density in this image is 0.87 mJy/beam.
 }
\label{smoking}
\end{center}
\end{figure}

Figure \ref{smoking} shows two diffuse clouds of radio emission whose origin is unclear.  The most likely hypothesis is that they represent the remnants of a classical double-lobed radio galaxy in which the central engine has switched off, leaving a remnant radio galaxy.  We refer to this object as the ``Smoking Gun.''

A tentative identification of the host (marked A in Figure \ref{smoking}) is the galaxy WISEA J204837.65--491115.2, at a redshift of 0.10  \citep{bilicki16} and which is detected as a 260 \ujy\ unresolved source in the ASKAP image. At that redshift, the largest angular size across the lobes is 530 kpc,  which is not unusual for double-lobed radio galaxies.
An alternative identification (marked B in Figure \ref{smoking}) of the host is an isolated unresolved 200 \ujy\ radio source 
which appears to be coincident with the galaxy DES J204835.43-491137.5, at
z=0.937 \citep{zhou21}. If the ``Smoking Gun'' were actually at this redshift, then the source's large inferred radio size (2.25 Mpc) would make it a member of the rare class of “Giant Radio Galaxies" (see Section \ref{GRG}) and it would also be very luminous ($\sim 4 \times 10^{23}$ W/Hz). 
This seems unlikely for a fading remnant, and so we think A, at a redshift of 0.1,  is more likely to be the host galaxy.
We note that the northern lobe is unusually circular and resembles the ORCs (``Odd Radio Circles'') shown in Figure \ref{sample}, except  that the ORCs don't show a continuous rise of their
surface brightness towards their centers.

Such remnant radio galaxies have previously been reported \citep[e.g.][]{brienza17, mahatma18,  saripalli12},   but these new observations probe a lower level of surface brightness than  earlier studies. Another remnant radio galaxy imaged by ASKAP has also been recently reported \citep{quici21}. However, none of the previously reported remnant radio galaxies has a circular lobe resembling that in Figure \ref{smoking}.

\begin{figure}
\begin{center}
\includegraphics[width=8cm, angle=0]{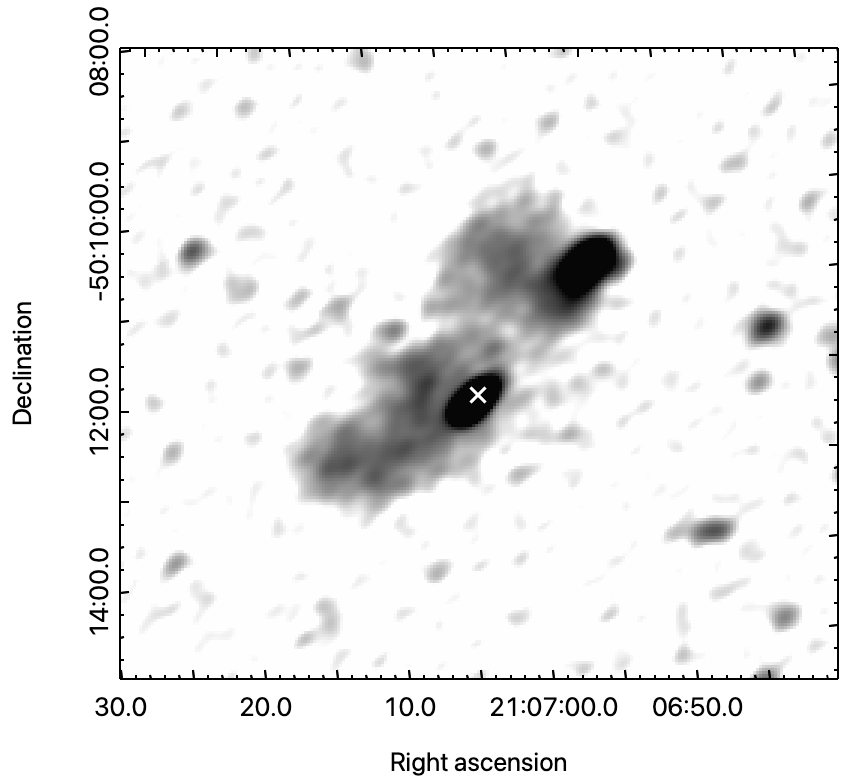}
\caption{
EMU PS J210700.0--501128.8 is an ambiguous case, appearing at first to be a double-lobed source with material blown to the east. But with no host between the bright patches, and the southern bright component coincident with a quasar, marked with an ``X'',  these may be two independent sources with serendipitously similar appearances The peak flux density in this image is 11.6 mJy/beam.
 }
\label{fig:sideways2}
\end{center}
\end{figure}

Figure \ref{fig:sideways2} shows the radio source EMU PS J210700.0$-$501128 (also detected as SUMSS J210704$-$501206). It appears similar in some ways to PKS~2130-538 shown above (Figure \ref{fig:ghosts}) with two bright patches and diffuse tails, presumably blown to the east by relative motion through an external medium. However, there is no obvious host galaxy between the lobes, only a scattering of faint DES galaxies.  Instead, the bright southern lobe is coincident with the quasar WISEA J210703.75$-$501207.7 at a redshift of 0.197 \citep{monroe16}, also detected in the 	Second ROSAT all-sky survey \citep{2016A&A...588A.103B}. This raises the possibility that the two ``lobes'' are two independent tailed radio galaxies with very similar morphologies.  
\rev{An apparent overdensity
of galaxies visible in both DES and WISE is embedded in the faint southern emission. However,  there is no cataloged cluster near this location, and we found photometric redshifts \citep{zou19} 
for  only 3 galaxies within the source confines, that would be consistent with the
quasar redshift, so there is no evidence for a cluster.  }

\subsection{Odd Radio Circles}

\begin{figure}
\begin{center}
\includegraphics[width=8cm,angle=0]{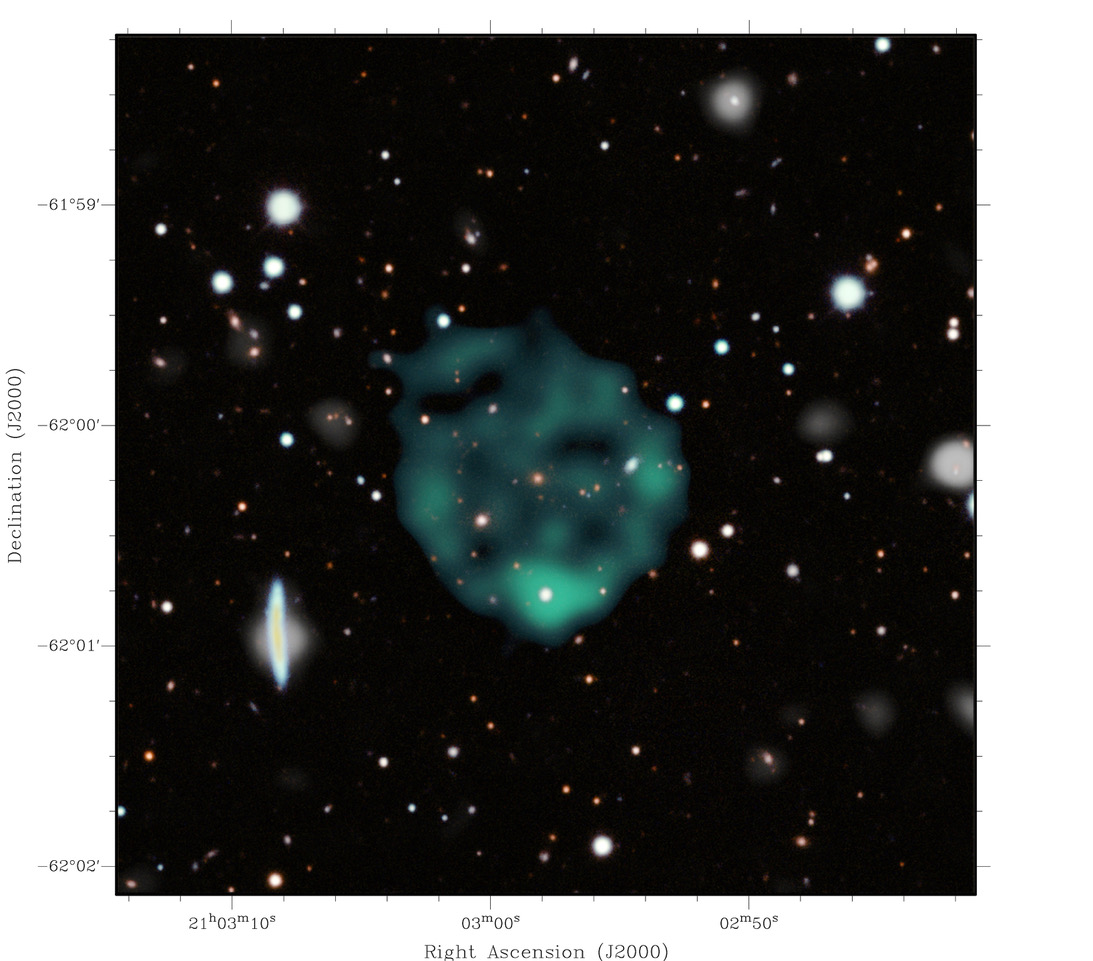}
\caption{An image of the first ``Odd Radio Circle'', or ORC, found in EMU-PS \citep{orc}. It has no optical counterpart to the diffuse ring, or to other diffuse structure, but has a galaxy at its centre which may be the origin of the ring. The image is based on EMU-PS data at native resolution, but enhanced to show faint features as described in \citet{orc}, particularly the internal structure or “spokes” of the ORC. Radio data are shown in green, and DES optical data are shown in turquoise, magenta, yellow and red, and mainly appear in this image as white.
}
\label{orc1}
\end{center}
\end{figure}

It has been predicted that, because EMU would  observe a previously inaccessible part of observational parameter space, it would probably make unexpected discoveries \citep{norris17b}. Nevertheless, we were surprised 
to find an apparently new class of object appearing in the EMU-PS, consisting of circles of radio emission, typically one arcmin across, with no optical or infrared counterpart \citep{orc}. The first of these Odd Radio Circles (ORCs) to be identified is shown in Figure \ref{orc1}, and another example is shown in Figure \ref{sample}. We do not yet understand the nature of these objects, nor whether they are a  single class of object or multiple classes. Since discovering them in EMU-PS, we have subsequently observed them with several other telescopes to confirm their reality, and have been able to rule out some potential explanations such as supernova remnants or starburst rings.  We have also found more examples in other ASKAP fields \citep[e.g.][]{koribalski21}. 
Several ORCs have a galaxy at the centre, typically at a redshift of $\sim$ 0.3 in the currently known examples. Potential explanations are that these central galaxies may be the origins of  spherical shock waves which we see in projection as a ring, or else that we are seeing  end-on radio lobes.

\subsection{Nearby Galaxies}
\rev{
The design of ASKAP was largely driven by  its two largest survey science projects: the EMU continuum survey \citep{emu} and the WALLABY spectral line survey (Koribalski 2012) with the latter
aiming to map neutral hydrogen (H\,{\sc i}) over the entire extragalactic sky in the declination range from $-90\degr$ to $+30\degr$ to a redshift of $\sim$0.26. WALLABY will generate H\,{\sc i} image cubes at $\sim$30 arcsec resolution and $\sim$1.6 mJy beam$^{-1}$ per 4~km\,s$^{-1}$ channel sensitivity and is expected to detect around half a million galaxies with a mean redshift of $\sim$0.05 \citep{koribalski20}. 

The relationship between the integrated radio continuum emission of star-forming galaxies, unattenuated by interstellar dust, and their star formation rate (SFR) has been extensively studied, \citep[e.g.][]{condon92,tabatabaei17,davies17}, and the broad correlation is well documented, \citep[e.g.][]{condon02,murphy09,murphy11,molnar21}. However, the detailed correlations and the underlying mechanisms are still the subject of much debate  \citep[e.g.][]{heesen14}.
Furthermore, the relationship between the atomic neutral hydrogen gas content of galaxies (for $z < 0.2$, where individual galaxy detections are feasible) and their star formation rate requires further investigation \citep[e.g.][]{wong16,bera19}.

Our most extensive knowledge of the southern sky in neutral hydrogen currently comes from the }
low-resolution H\,{\sc i} Parkes All Sky Survey \citep[HIPASS;][]{barnes01}  which covers the sky from $-90\degr$ to $+25\degr$. HIPASS produced a catalogue of $\sim$5000 galaxies out to a redshift of $z = 0.04$ \citep{koribalski04, meyer04}.  Corresponding 20-cm radio continuum maps (CHIPASS) were created by \citet{calabretta14}. ASKAP delivers a 90-fold (for EMU) or 30-fold (for WALLABY) improvement in angular resolution compared to the HIPASS single-dish beam of $\sim$15.5 arcmin. 

\rev{The survey characteristics of EMU and WALLABY imply that we expect nearly all of the $\sim$5000 catalogued HIPASS galaxies to be detected by EMU, and to be well resolved by both EMU and WALLABY. Therefore, the combination of EMU radio continuum and WALLABY H\,{\sc i} spectral line measurements of nearby galaxies, combined with other multi-wavelength data, offers the opportunity to study the relationships between star formation, radio continuum emission, and H\,{\sc i} emission in great detail. Here we start to explore this field using the EMU-PS observations of a small sample of nearby galaxies which have been detected in HIPASS. }

\begin{figure*} 
    \centering
    \includegraphics[width=17cm]{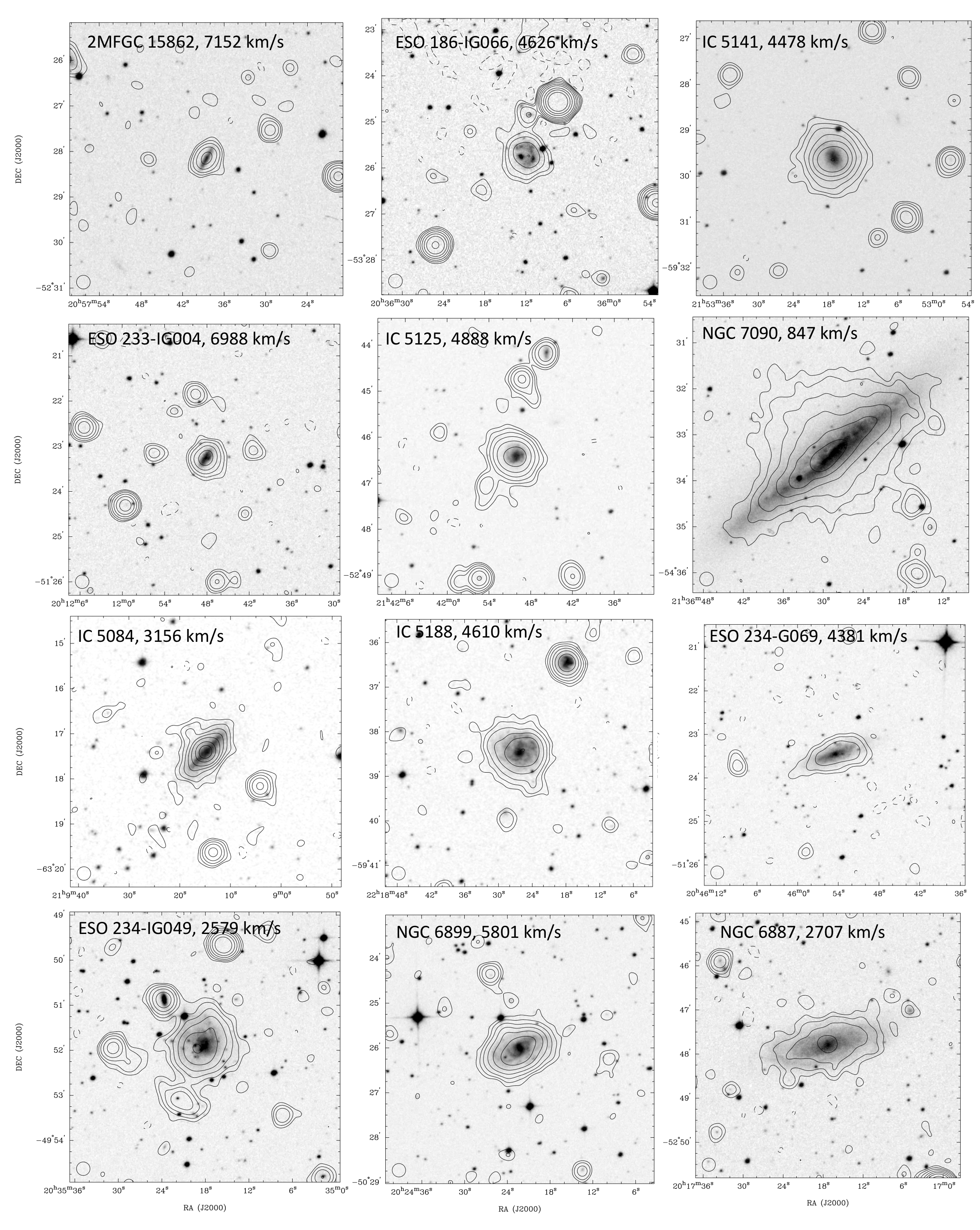}
    \caption{A selection of ASKAP-detected nearby galaxies in the EMU-PS. Optical DSS2 $R$-band images are overlaid with ASKAP radio continuum contours. The contour levels are --0.09, 0.09 ($\sim$3$\sigma$), 0.18, 0.36, 0.75, 1.5, 3.0, 7.5, 15.0, and 30 mJy\,beam$^{-1}$. The convolved 18 arcsec beam is shown in the bottom left corner of each panel. The galaxy name and heliocentric velocity (all but one, ESO233-IG004, from HIPASS) are also displayed. The velocity of ESO233-IG004 is taken from \citet{jones2009}. }
    \label{fig:nearbygalaxies}
\end{figure*}

\begin{figure} 
    \centering
    \includegraphics[width=8cm]{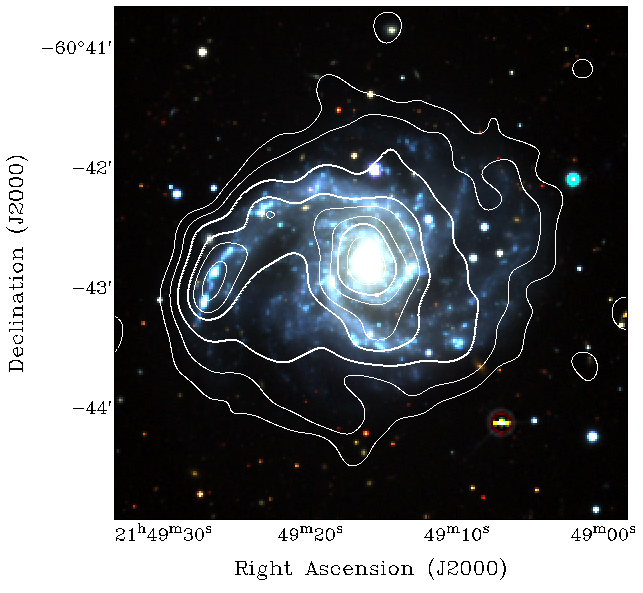}
    \caption{
    DES-DR1 optical composite image of the nearby face-on spiral galaxy NGC~7125 overlaid with contours from the EMU-PS. The contour levels are: 0.1, 0.25, 0.5, 1.0, 1.3, 1.6, 2.0 and 2.4 mJy\,beam$^{-1}$. NGC~7125 and its neighbour NGC~7216 form an interacting galaxy pair (HIPASS J2149--60) with a large pool of hydrogen gas for star formation.   }
    \label{fig:ngc7125}
\end{figure}

Of the 89 catalogued HIPASS sources in the EMU-PS area, 63 are clearly detected by EMU-PS in the radio continuum. We expect most of the remaining HIPASS sources to be detected in the full sensitivity main EMU survey. A selection of these galaxies is shown in Figs.~\ref{fig:nearbygalaxies} and \ref{fig:ngc7125}. While no WALLABY H\,{\sc i} data currently exist for this field, high-resolution ATCA H\,{\sc i} images are available for some of galaxies.

\rev{

The most interesting galaxy in Fig.~\ref{fig:nearbygalaxies} is the edge-on spiral NGC~7090. Using the ATCA, \citet{heesen16, heesen18} obtained detailed radio continuum maps, finding a radio halo with polarised emission up to 6 kpc above the disk correlating with extraplanar H$\alpha$ emission. ATCA H\,{\sc i} images \citep{dahlem05} reveal an asymmetric, slightly disturbed disk matching the stellar extent. Another galaxy of interest is the nearly face-on spiral NGC~7125 (see Fig.~\ref{fig:ngc7125}), which forms an interacting pair with its northern companion NGC~7126, separated by six arcmin ($\sim$80 kpc). ATCA H\,{\sc i} maps \citep{nordgren97}
show a large gas envelope encompassing both galaxies. Once detailed H\,{\sc i} spectral line and radio continuum maps are available for large numbers of nearby resolved galaxies, the local and global star formation rates and efficiencies can be analysed as a function of H\,{\sc i} column density and environment \citep[e.g.][]{koribalski09, wong16}. 
}

\subsection{Giant Radio Galaxies}
\label{GRG}

Giant Radio Galaxies (GRGs) were originally defined as Radio Galaxies (RGs) whose projected linear size was greater than 1~Mpc for a Hubble constant of H$_0$=50\,km\,s$^{-1}$\,Mpc$^{-1}$
\citep[e.g.][]{ishwarachandra99}. However, based on the currently accepted value of H$_0 \sim $70\,km\,s$^{-1}$\,Mpc$^{-1}$, RGs larger than 0.7~Mpc are now also considered GRGs.
In the compilation of GRGs by \citet{kuzmicz18}, the EMU-PS area contains only a single GRG, namely PKS 2014$-$558, first mentioned as a GRG by \citet{jones92} and recently studied in detail by \citet{cotton20}.

In a recent ASKAP observation covering 30 \sqdeg\ centered on the Abell 3391/3395 galaxy cluster pair, and of comparable depth and angular resolution as the EMU-PS, \citet{bruggen21} found the surface density of GRGs $\gtrsim$1\,Mpc to be $\sim0.8$\,deg$^{-2}$, and that of GRGs $\gtrsim0.7$\,Mpc to be at least $\sim1.7$\,deg$^{-2}$, suggesting that the EMU-PS should contain $\sim$200 and $\sim$460
such GRGs, respectively.

From a preliminary visual inspection of the EMU-PS area, biased towards sources of larger angular size and featuring a radio nucleus, we found $\sim$120 GRGs larger than 1~Mpc and a similar number with sizes between 0.7 and 1~Mpc. 
We visually cross-identified these with the Dark Energy Survey images and catalogues \citep{DES}, and estimated linear sizes based on photometric redshifts \citep{bilicki16,drlicawagner18, zou19}. The number of GRGs in EMU-PS is likely to increase with a more thorough visual inspection, the results of which will be reported by Andernach et al. (in prep.).

\subsubsection{The Giant Radio Galaxy EMU PS J205139.8--570434 }
\begin{figure*}
\begin{center}
\includegraphics[width=20cm, angle=0]{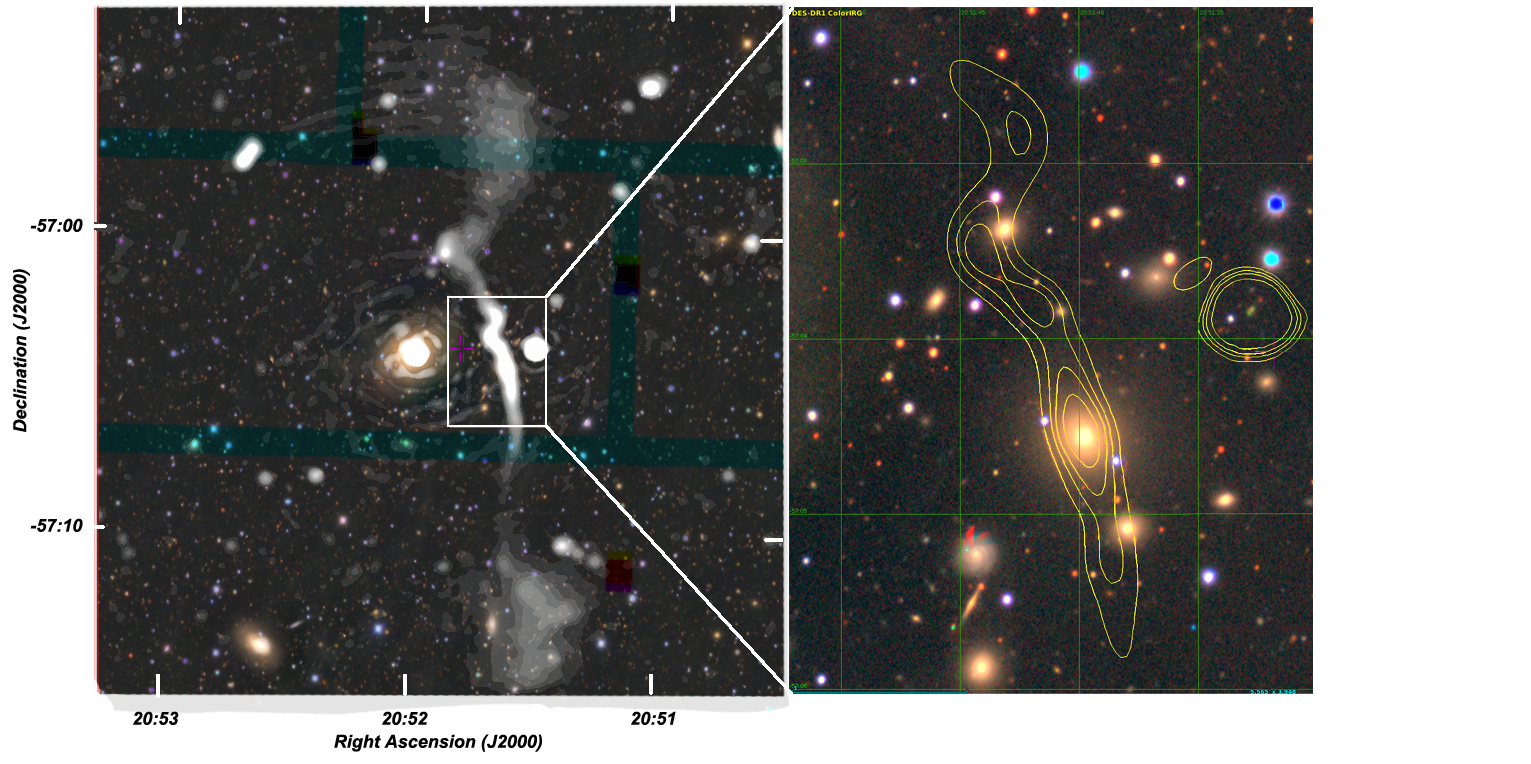}
\caption{(Left) The Giant Radio Galaxy (GRG) EMU PS J205139.8--570434, with radio (at native resolution) shown in greyscale, overlaid on the DES DR1 colour image. The GRG consists of the roughly north-south jet and the two diffuse plumes above and below it. The strong source to the east, surrounded by diffraction rings, is the well-studied galaxy IC\,5063. (Right) A contour diagram of the central part of the GRG at 18 arcsec resolution, overlaid on the DES DR1 colour image. Contour levels are  3, 7, 12 and 18 mJy/beam.
 }
\label{GRG1}
\end{center}
\end{figure*}

The GRG EMU PS J205139.8--570434 (hereafter GRG~J2051--5704), shown in Figure \ref{GRG1}, is hosted by
2MASX~J20513976--5704334 at $z_{sp}$=0.0602 \citep{jones2009}
and its radio emission can be traced over a Largest Angular Size $\sim22.1$\arcmin\ and thus a linear projected size of 1.53\,Mpc. It has an FR\,I radio morphology  whose  jets are oriented roughly North-South, feature several wiggles, and
terminate in diffuse lobes at both extremes of the source. 
The strong (S$_{944}$=2.12\,Jy) source 3.1~arcmin due E, surrounded by 
diffraction rings, is the well-studied galaxy IC\,5063 at $z_{sp}$=0.01135.

The central region of the new GRG  can be recognized in 
the SUMSS \citep{bock99} images and was even detected at mm wavelengths by the South Pole Telescope
\citep{mocanu13}. However, the full extended structure shown here has not been previously detected, despite extensive imaging of the 
neighboring source IC\,5063  (\citealt{murphy10} and references therein), probably because of the brightness sensitivity and dynamic range limitations.

The host galaxy of GRG J2051--5704 is the brightest galaxy of cluster
400d~J2051--5704 \citep[aka MCXC~J2051.6--5704,][]{burenin07,piffaretti11}
at a redshift of 0.0599. The radio morphology is reminiscent of
archetypal FR\,I sources such as 3C\,31 and Hydra\,A, with wiggles consistent with either
the presence of jet instabilities due to interaction with the surrounding
gas, or jet precession \citep[e.g.][]{nawaz16}. 
While FR\,I type sources are rare among GRGs larger than 1~Mpc, a recent
LOFAR image showed that the GRG 3C\,31 also had an extent $>$~1~Mpc \citep{heesen18}.

It is likely that more of these very large GRGs will be detected 
with 
\rev{next-generation radio telescopes such as ASKAP}
thanks to the combination of good angular resolution \rev{necessary to 
reveal the inner jet structure and identify the host}, as well as high sensitivity to 
the low surface brightness features such as the outer tails \rev{or lobes. We note that \citet{turner18} showed from simulations that more sensitive, or lower frequency, observations will reveal FR\,I galaxies to be much larger than previously thought}.

\subsection{Radio counterparts to 6dF Galaxies}

EMU and the EMU-PS overlap the 6dF Galaxy Survey \citep[6dFGS; ][]{jones2004,jones2009}, a spectroscopic survey of most of the southern sky containing 125,071 galaxy redshifts with a median redshift of $0.053$. While a variety of selection criteria were used for 6dFGS, most 6dFGS galaxies are brighter than $K=12.65$ and have redshifts of $z<0.15$. In the EMU-PS region, there are 2506 6dFGS galaxies and, as we discuss below, a large fraction of these galaxies are detected by the EMU-PS.

We measured the flux density of each 6dFGS galaxy using the pixel in the radio continuum maps corresponding to each galaxy’s position. \rev{This will underestimate the total flux density of spatially resolved galaxies, and aperture bias is relevant as the $13^{\prime\prime}\times 11^{\prime\prime}$ EMU-PS beam and $6.7^{\prime\prime}$ 6dFGS spectroscopic fibre correspond to less than $3.4~{\rm kpc}$ and $2.0~{\rm kpc}$ respectively for galaxies within $60~{\rm Mpc}$ of Earth. Despite these limitations, our preliminary measurements allow us to quantify the fraction of 6dFGS galaxies that are radio sources and allows us to push fainter than blind radio source catalogues. Of the 2506 6dFGS galaxies in the EMU-PS region, 1887 (75\%) have a flux density greater than $75$\ujybm, corresponding to $\gtrsim 3\sigma$.} As star-forming and passive galaxies have different distributions of radio continuum luminosities, we roughly split these two populations using the presence and absence of ${\rm H\alpha}$, measured from the 6dFGS spectra \citep{jones2004,jones2009}. To quantify noise and source confusion, we also measure flux density at positions offset by 100 pixels (200 arcsec, so well outside the relevant galaxy).

\begin{figure}
    \begin{center}
    \includegraphics[width=9cm, angle=0]{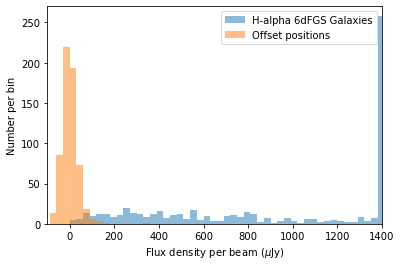}
    \caption{The histogram of EMU-PS flux densities for 6dFGS galaxies with ${\rm H\alpha}$ emission and $K<12.65$, along with the corresponding histogram of flux densities measured at offset positions. Just 17 of the 623 6dFGS galaxies with detectable ${\rm H\alpha}$ emission have radio flux densities below $75$\ujybm. 
    }
    \label{fig:sf_hist}
    \end{center}
\end{figure}

In Figure~\ref{fig:sf_hist} we present the histogram of EMU-PS flux densities of 6dFGS galaxies with ${\rm H\alpha}$ and with $K<12.65$, along with the histogram of flux densities measured at offset positions. Roughly half of the star forming galaxies are fainter than $1~{\rm mJy}$ and would not have  been detected by previous generations of wide-field radio continuum surveys. Almost all 6dFGS galaxies with detectable ${\rm H\alpha}$ emission and $K<12.65$ in the EMU-PS area are detected, with just 17 of the 623 galaxies having flux densities below  $75~{\rm \mu Jy}$. \rev{For comparison, at the 623 offset positions there are just 22 flux density measurements brighter than $75$\ujybm~ (corresponding to $\gtrsim 3\sigma$) and only 8 flux density measurements brighter than $125$\ujybm~ (corresponding to $\gtrsim 5 \sigma$).}

\begin{figure}
    \begin{center}
    \includegraphics[width=9cm, angle=0]{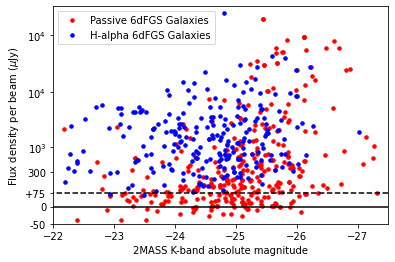}
    \caption{The EMU-PS flux densities of $K<12$ galaxies with and without ${\rm H\alpha}$ emission as a function of absolute magnitude. \rev{The dashed line shows $75~{\rm \mu Jy}$, roughly corresponding to $3\sigma$.} Almost all star forming galaxies are detected by the EMU-PS. While low mass passive galaxies can have no detectable radio continuum emission, all but one $M_K<-26$ passive galaxy has a positive radio continuum flux density.
    }
    \label{fig:flux_absmag}
    \end{center}
\end{figure}

The radio continuum flux densities of $K<12$ galaxies with and without ${\rm H\alpha}$ emission is presented in Figure~\ref{fig:flux_absmag}. Most $K<12$ star-forming galaxies are detected by EMU-PS, and a significant fraction of passive galaxies are also detected. While the lowest mass passive galaxies are often undetected by EMU-PS, all but one of the $M_K<-26$ passive galaxies has a positive radio continuum flux density, presumably resulting from AGNs. This is consistent with \citet{brown2011} and \citet{sabater2019}, who have concluded that all massive elliptical galaxies are radio continuum sources, using NVSS and LOFAR respectively. When complete, EMU will detect thousands of nearby $M_K<-26$ elliptical galaxies, enabling characterisation of the radio luminosities of these objects and the AGNs they host. 

\subsection{Comparison with Gaia}

\begin{figure}
    \begin{center}
    \includegraphics[width=9cm, angle=0]{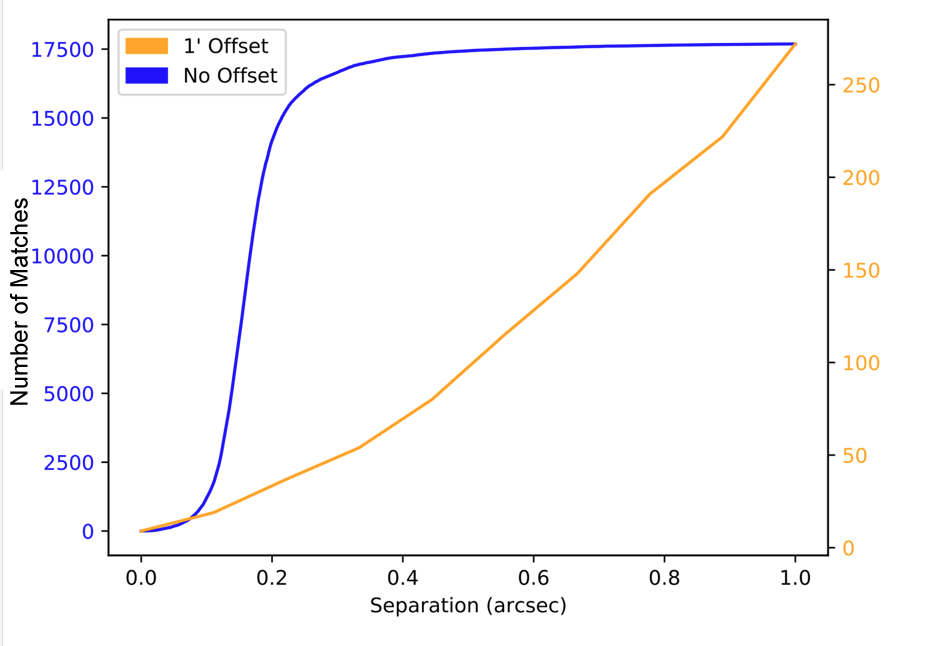}
    \caption{The number of cross-matches between DES counterparts to EMU-PS sources, and Gaia sources,  for unshifted data (blue), and data shifted by one arcmin (orange).}
    \label{fig:gaia-xmatch}
    \end{center}
\end{figure}

The Gaia project \citep{gaia} has measured the parallax and proper motion of over a billion stars, and as a by-product has also identified a number of quasars and compact galaxies \citep{bailer}. To produce a catalogue of candidate radio-loud quasars, we therefore cross-match the DES counterparts to EMU radio sources against the Gaia EDR3 catalogue \citep{gaia21}. We use the same technique as in earlier cross-matches in this paper, resulting in a plot of cross-matches as a function of search radius for shifted and unshifted data, shown in Figure \ref{fig:gaia-xmatch}. As a result of this test, 
we choose a cross-match radius of 0.2 arcsec, resulting in 14,174 cross-matches in the unshifted data, and 14 cross-matches in the shifted data, indicating a false-ID rate of 0.1\%. 

\begin{figure}
    \begin{center}
    \includegraphics[width=8.5cm, angle=0]{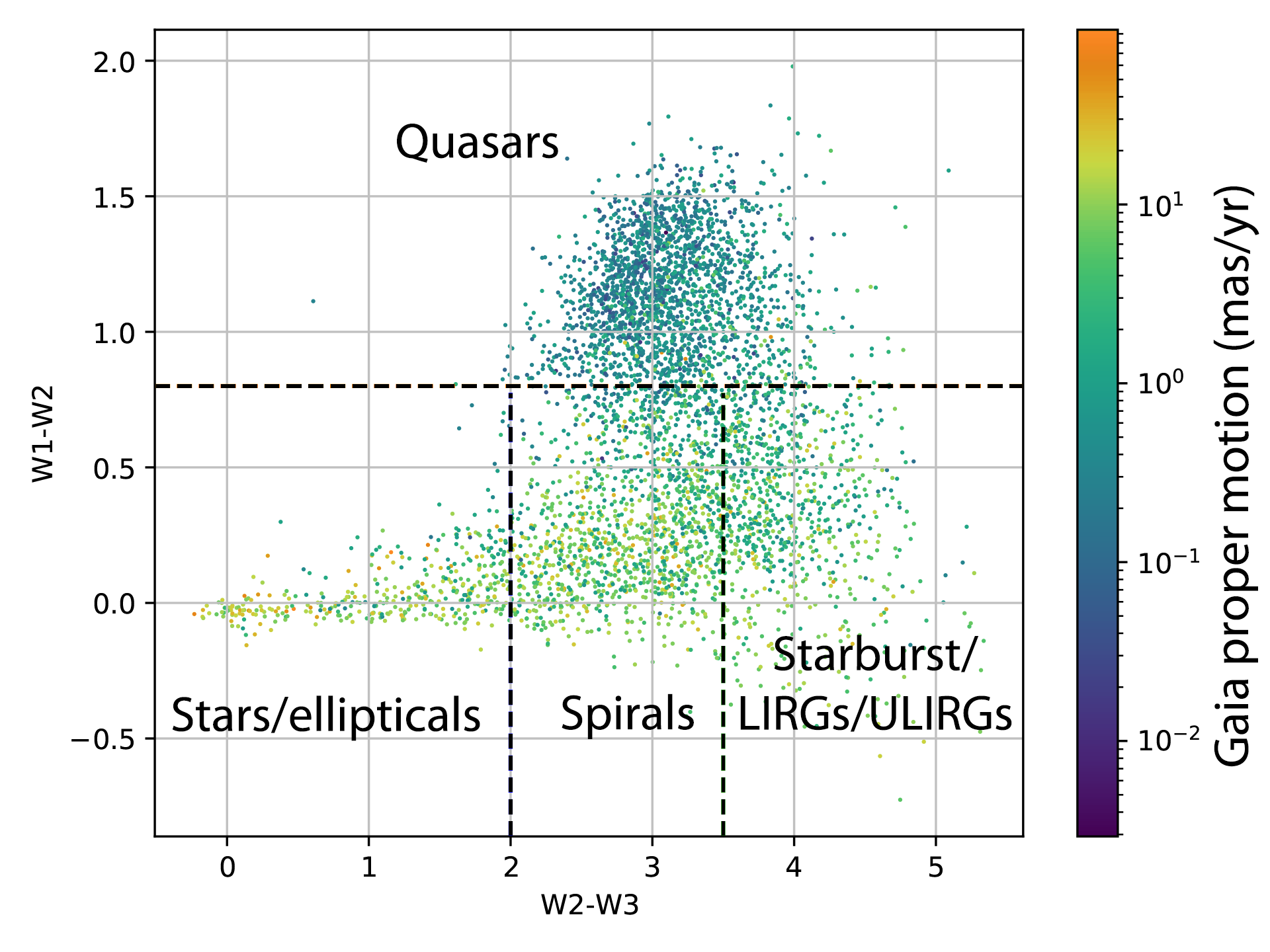}
    \caption{The AllWISE colour-colour plot for EMU-PS sources cross-matched with Gaia, colour-coded according to their measured proper motion. The dashed lines divide the graph into the regions identified by   \citet{jarrett17}.
    }
    \label{fig:gaia-wise}
    \end{center}
\end{figure}

To explore the infrared properties of this sample of Gaia-selected sources, \rev{we need to use the WISE W3 band, which is absent from the  CWISE catalog, and so we must match our sources against the AllWISE catalog \citep{cutri14}. We therefore } cross-match the list of 14,174 sources, using a search radius of 1 arcsec, against the AllWISE catalogue, resulting in a catalogue of 11,142 sources with WISE W1, W2, and W3 flux densities. These sources are shown in Figure \ref{fig:gaia-wise}, colour-coded by their proper motion. 

The sources with the lowest proper motions lie in the region identified by \citet{jarrett17} as being dominated by quasars, while a higher level of proper motion is seen in the region dominated by galaxies. This effect was also noted by \cite{bailer} who explained it as extended galaxies not having well-defined centroids, causing the measured Gaia position to vary, resulting in an apparent proper motion. The group with the highest proper motion (coded as yellow) lie in the region designated as stars, confirming that these DES sources are indeed stars. However,  most, if not all, of these  ``stars''  are probably false IDs and do not correspond to radio sources. 

Of the 11,142 galaxies with W1, W2, W3 flux densities, 2604 have W1-W2 $>$0.8, and we refer to these as quasar candidates. To estimate the false-ID rate, we repeated the above selection process after shifting the declination by 1 arcmin, and this resulted in 2312 sources with W1, W2, W3 flux densities, of which 503 have W1-W2 $>$0.8. We therefore expect that 81\% of our quasar candidates are radio-loud quasars, assuming that no other types of source fall in that part of the WISE colour diagram. 

The combination of the Gaia selection and the WISE colour selection has therefore yielded a catalogue of 2312 radio-loud quasar candidates, of which about 81\% are true radio-loud quasars. Using the same technique on the entire EMU survey will yield a catalogue of about 230,000 radio-loud quasar candidates, representing a significant increase in the number of known radio-loud quasars.

\subsection{Clusters of Galaxies}

 Galaxy clusters represent some of the largest gravitationally-bound structures in the Universe, and radio emission provides an insight into their formation and evolution. 
They evolve and grow  through a variety of processes including passive accretion of gas, consumption of small galaxy groups, and violent merger events which can deposit vast amounts of energy \citep[$\sim10^{64}$ erg, e.g.][]{ferrari2008} into the intracluster medium (ICM).

Many merging galaxy clusters host vast and  enigmatic radio continuum sources. These diffuse radio sources are broadly classified into two categories: radio relics (or cluster radio shocks) and radio halos \citep[see][for a recent review]{vanweeren2019}. To date, some $\sim70$ clusters are known to host radio relics, and some $\sim65$ clusters are known to host radio halos\footnote{An up-to-date database of these sources is maintained at \url{GalaxyClusters.com}.}.

Radio relics are highly-extended (typically $\sim1$\,Mpc), highly-polarised, diffuse synchrotron sources that lie toward the periphery of galaxy clusters. They often exhibit curved  morphologies and filamentary sub-structures and are thought to be powered by shocks which generate relativistic electrons through a form of diffusive shock acceleration.

Radio halos, on the other hand, are largely amorphous, unpolarised diffuse synchrotron sources that are centrally-located in merging galaxy clusters, and roughly follow the distribution of the thermal plasma in the ICM (as traced by X-ray emission).%
The most commonly-accepted scenario is that radio halos are powered by turbulence injected in the ICM during cluster merger events, although there are alternatives based on collisions between cosmic ray protons (CRp) and thermal protons from the ICM \citep[for a review, see][]{BrunettiJones2014}. On smaller scales ($\lesssim0.5$\,Mpc), ``mini-halos" are relatively small diffuse radio sources that are generally co-located with powerful, radio-loud, brightest cluster galaxies in relaxed clusters. One theoretical explanation for mini-halos is that `core-sloshing' in the ICM from minor or off-axis mergers produces small-scale turbulence which can then provide sufficient energy for re-acceleration of the relativistic electrons.  

For relics, halos, and mini-halos, the shape of the synchrotron emitting spectrum provides a diagnostic for these relativistic particle (re-)acceleration processes.
Historically, our understanding of the relevant physics has been limited by (i) the generally relatively poor quality of low-frequency radio data, (ii) missing short spacings, leading to loss of highly-extended radio emission, and (iii) narrow bandwidths, limiting the spectral shape measurements. By circumventing these limitations with the EMU survey, the sample of clusters suitable for study can be increased by at least two orders of magnitude. 

There are already $\sim$20 known X-ray detected galaxy clusters in the EMU-PS area. All show compact or moderately-extended radio sources that are likely associated with AGN. The eROSITA survey \citep{2021A&A...647A...1P} will provide many more X-ray clusters for radio investigation. The first eROSITA all-sky survey (eRASS1) will find as many as 1.5 clusters/\sqdeg; more than 3 clusters/\sqdeg are expected after all eight all-sky surveys are completed \citep[e.g.,][]{2018MNRAS.481..613P}. Hence, we expect as many as 400 X-ray-detected clusters will soon be available in the EMU-PS area alone, and more than 10,000 AGN.

Early EMU/eROSITA results on the  Abell 3391/95 galaxy cluster system
\citep{2021A&A...647A...2R,bruggen21} have already helped constrain physical processes in the merger.  We show two additional examples of diffuse radio sources detected in the EMU-PS area in Figure~\ref{fig:spt2023} and Figure~\ref{fig:spt2032}. 

Other cluster catalogs will similarly provide important targets for EMU-PS and the full EMU Survey.  \rev{For example,} \citet{aguena21}  provide the WaZP catalog of 60542 clusters from the DES \footnote{\url{https://www.linea.gov.br/catalogs/wazp/}}. About 30 of their Brightest Cluster Galaxies (BCGs) are coincident with extended RGs in the EMU-PS, and will be discussed by Andernach et al. (in preparation).

\subsubsection{SPT-CL J2023-5535}
\begin{figure}
    \begin{center}
    \includegraphics[width=0.48\textwidth]{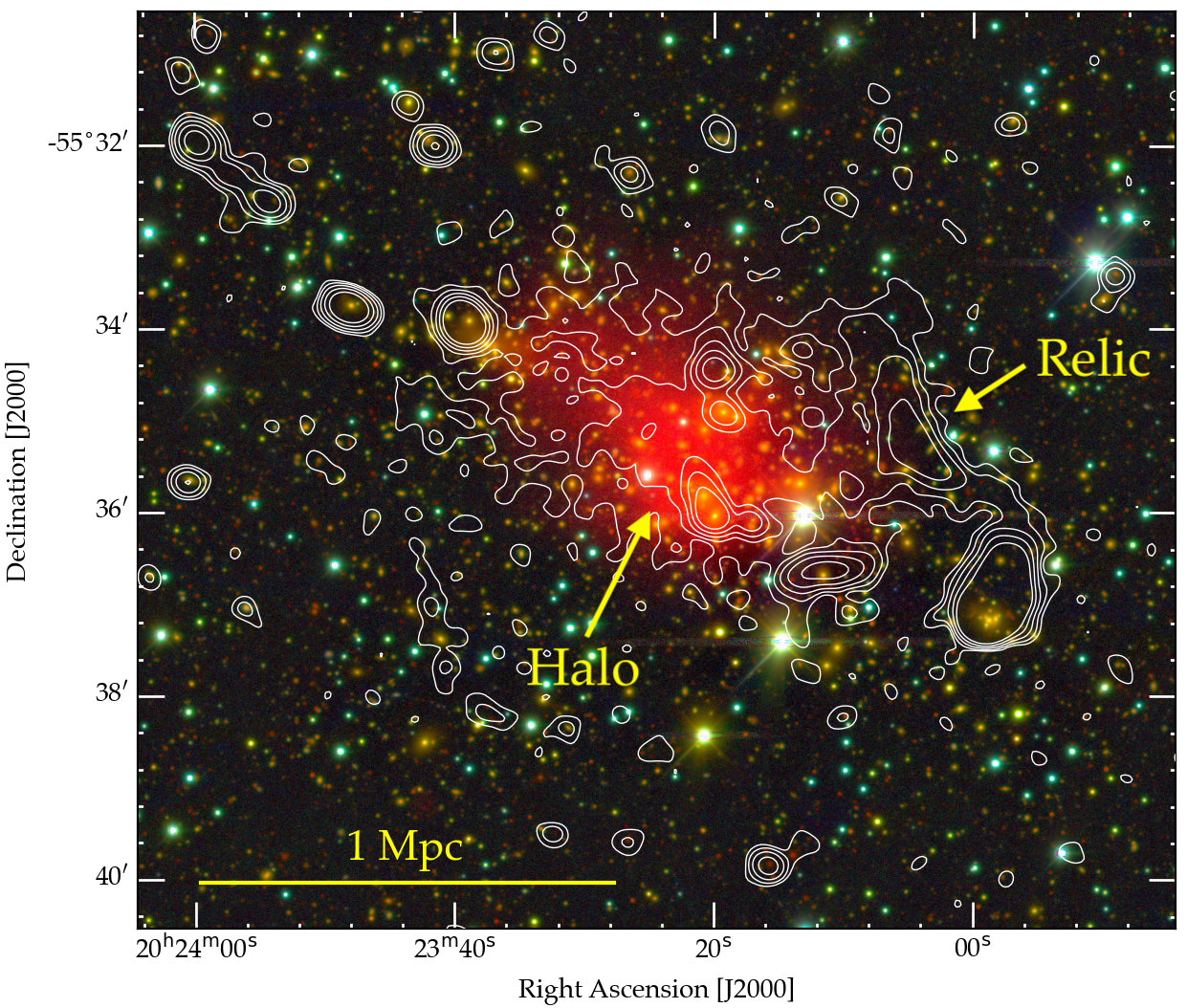}
    \caption{Multi-wavelength composite image of SPT-CL~J2023$-$5535.  Contours denote the EMU-PS surface brightness at 944\,MHz at 18\arcsec\ resolution, at $3\sigma_{\rm{rms}} \times 2^{n}$ where $n=0, 1, 2, 3, 4$ and $\sigma_{rms}=\sim25$\ujybm. Background colormap shows a composite {\it g}, {\it r}, and {\it i} image from DECam. X-ray emission from {\it Chandra} is also overlaid in red. New diffuse radio sources identified by \protect\cite{hyeonghan2020} are also indicated.}
    \label{fig:spt2023}
    \end{center}
\end{figure}

Figure~\ref{fig:spt2023} presents the radio halo and relic in the massive merging galaxy cluster, SPT-CL~J2023$-$5535 ($z=0.23$), reported by \cite{hyeonghan2020}. Their weak-lensing analysis has revealed significant sub-structure in this massive ($M_{200}=1.04\pm0.36\times 10^{15}~M_{\sun}$) cluster, which comprises three sub-clusters.

The merger event between the Eastern and Central sub-clusters appears to have generated a $\sim0.5$\,Mpc radio relic on the Western edge of the central sub-cluster. The results presented by \cite{hyeonghan2020} show an unusually flat spectral index $\alpha_{\rm{int}}=-0.76 \pm 0.06$, which may indicate that this relic is powered by the re-acceleration of fossil electrons, perhaps originally seeded by a nearby (photometric) cluster-member AGN. Follow-up observations at other radio frequencies will be required to confirm this flat spectrum.

\subsubsection{SPT-CL J2032-5627}
\begin{figure}
    \begin{center}
    \includegraphics[width=0.48\textwidth]{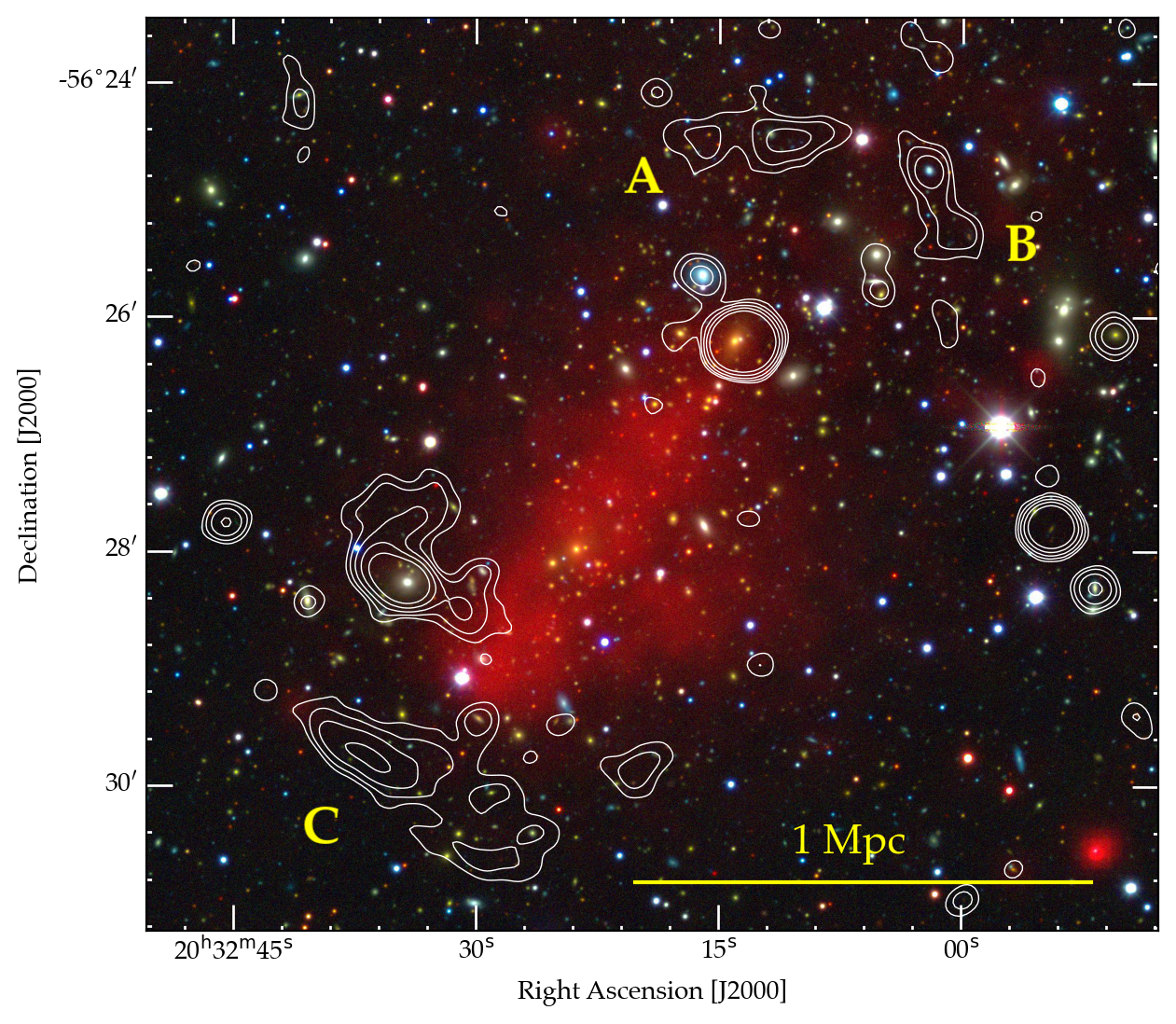}
    \caption{Multi-wavelength composite image of the cluster SPT-CL~J2032$-$5627. Colour map and contours are the same as Figure~\ref{fig:spt2023}, but with $25$~ks XMM-{\it Newton} surface brightness shown in red. It appears that the cluster hosts one of the rare class of double radio relics with the Northern (A and B) and Southern (C) relics as indicated.}
    \label{fig:spt2032}
    \end{center}
\end{figure}

Figure~\ref{fig:spt2032} shows a rare class of a possible double radio relic with an elongation of X-ray emission in the massive \citep[$M_{500}=4.77^{+0.71}_{-0.63}\times 10^{14}~M_{\sun}$;][]{Bulbul2019} cluster SPT-CL J2032$-$5627 ($z=0.28$). The North-Western (sources A \& B) and South-Eastern (source C) all exhibit steep radio spectra ($\alpha_{\rm{int}} = -1.75$, $\alpha_{\rm{int}} = -1.69$, and $\alpha_{\rm{int}} =-1.46$, respectively).

The highly asymmetric X-ray surface brightness profile and large projected separation between the radio relics in this cluster suggest that the merger event is occurring close to the plane of the sky. Curiously, no evidence of a shock has been found in the X-ray surface brightness. However, the presence of a cold front toward the leading edge to the South-East of the cluster may suggest that the lack of a shock detection is due to the relatively shallow depth ($25$~ks) of the existing XMM-{\it Newton} observations. See also \cite{Duchesne2021} for further discussion of SPT-CL~J2032$-$5627.

\subsubsection{A Mini-Halo in a poor cluster}
\begin{figure}
 \begin{center}
    \includegraphics[width=0.48\textwidth]{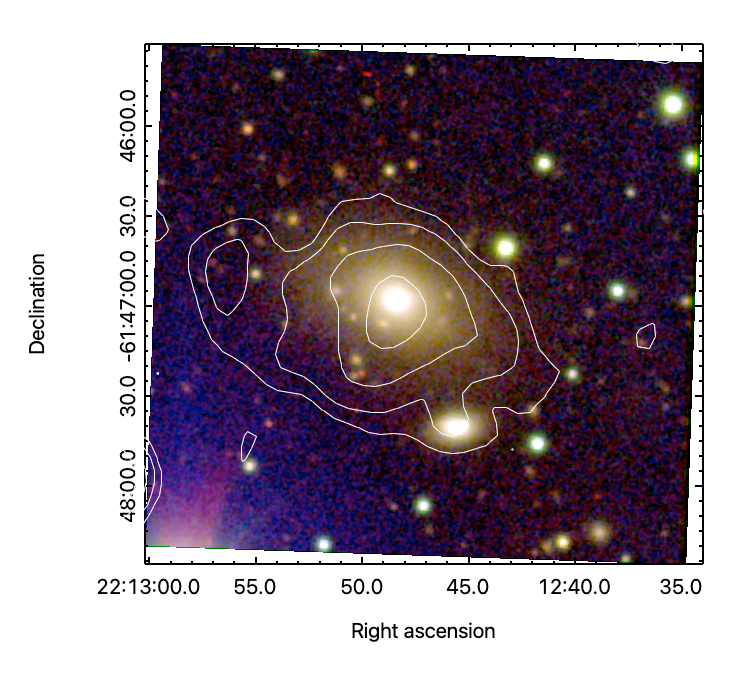}
    \caption{Radio contours overlaid on a multi-wavelength (irg) composite image of 6dFGS g2212485-614658 from DES. The radio image was made from the EMU-PS native resolution data by subtracting four unresolved sources and then convolving to a resolution of 18 arcsec.
Contours are at 75, 150, 300, and 600 \ujybm. The subtracted sources were at 22:12:48.64 $-$61:46:58.5 (14.9 mJy), 22:12:43.04 $-$61:46:50.1 (0.3 mJy), 22:12:51.19 $-$61:46:15.5 (0.2 mJy), 22:12:37.17 $-$61:47:24.5 (0.1 mJy)
    }
    \label{fig:minih}
    \end{center}
\end{figure}
 Figure \ref{fig:minih} shows a very low surface brightness structure, much fainter (50$\mu$Jy/beam) than the embedded compact 14.8$\pm$0.07 mJy source associated with 6dFGS g2212485-614658 at a redshift of z=0.054. The low surface brightness emission was found using the multi-resolution filtering technique of \cite{2002PASP..114..427R}, using a filter box size of 34\arcsec\ which removes the emission from compact components. It has a total extent of 150\arcsec\ (160~kpc) and a total flux density of 5$\pm$0.5 mJy, corresponding to a luminosity of $\approx 3\times10^{22}$ W/Hz.  No substructure is apparent.
 
 The diffuse structure could be the dying remains of a radio galaxy, faded to a luminosity comparable to the faintest AGN or to typical star-forming galaxies \citep{2007MNRAS.375..931M}.  The lack of radio structure, however, suggests that it could also be an underluminous mini-halo, an option we briefly explore here. 
 
 There is no cataloged cluster associated with 6dFGS g2212485-614658, although there are 5 galaxies with a similar redshift listed in Vizier, out to a separation of 17\arcmin\, (1~Mpc). This, and the presence of many smaller galaxies embedded in the 6dFGS g2212485-614658 envelope (Figure \ref{fig:minih}) suggest that this could be a poor cluster or group.
 
 The radio and extended (as opposed to AGN) X-ray luminosities of mini-halo systems are well-correlated \citep{2019ApJ...880...70G}. For this system, we determined upper limits to the bolometric (0.2 - 2 keV) X-ray emission using both RASS and XMM Slew archives, yielding limits in the range 1.4-2.4$\times10^{-13}$ erg/s/cm$^2$.
 At mid-range, the inferred luminosity upper limit of $10^{42}$ erg/s is two orders of magnitude lower than that of the mini-halo clusters summarized  in \citet{2019ApJ...880...70G} and than the value expected from the radio-X-ray correlation. 
 
 Probing mini-halo-like structures in these  poor-cluster, low X-ray luminosity type systems is important for understanding the physical mechanisms which form and continue to power the radio emission. We do not know whether the current observed radio-X-ray correlation is influenced by X-ray selection effects, or whether the correlation breaks down at very low cluster masses. 
 
 The sensitivity of EMU to very low surface brightness emission such as presented here will provide a powerful tool for exploring the connection between compact and extended AGN emissions and pure cluster/group particle acceleration processes in mini-halos.

\subsection{Cosmology}

The spatial distribution of radio sources is a tracer of the underlying matter distribution, and can be used to probe the formation conditions of radio galaxies, as well as the underlying fundamental ingredients and physics of the Universe. As these continuum sources are not easy to localise in redshift, we use measurements of angular clustering for the EMU-PS.

Here we use the \citet{1993ApJ...412...64L} estimator, which is defined as,
\begin{equation}
    w_{\mathrm{LS}}(\theta) = \frac{DD(\theta) + RR(\theta) - 2DR(\theta)}{RR(\theta)}
\end{equation}
where $DD(\theta)$ is the number of observed galaxy pairs at distance between $\theta$ and $\theta+d\theta$, $RR(\theta)$ is the number of random galaxies pairs at this separation, and $DR(\theta)$ is the number of observed-random pairs. We apply this statistic to the pilot survey catalogue, using the island data catalogue as the data vector ($D$), and  generated  random catalogues ($R$), normalising the number over all angles such the angular correlation function $w(\theta)$ functions as a probability excess or decrement relative to an entirely random distribution of galaxies on the sky.

The random catalogue ($R$) is generated using the method used in \citet{hale2018,Siewert2020} where random positions for simulated sources are generated across the EMU-PS field of view and for each simulated source a flux density is randomly assigned to the source using flux densities from the SKADS simulation \citep{wilman08}\footnote{We use the SKADS 1.4 GHz flux scaled to 944 MHz assuming a spectral index of -0.8. We also apply a minimum flux density cut on the 944 MHz converted SKADS flux of $\sim30 \mu$Jy/beam}. We assign noise to the flux density of the simulated source by sampling from a Gaussian distribution with spread given by the RMS at the random source location. A source remains within the random catalogue provided the simulated source peak flux density (where we assume the random sources are unresolved) added to the noise would be detectable at $\geq$5 $\times$ the RMS at the simulated source position.

\begin{figure}
    \begin{center}
    \includegraphics[width=8cm]{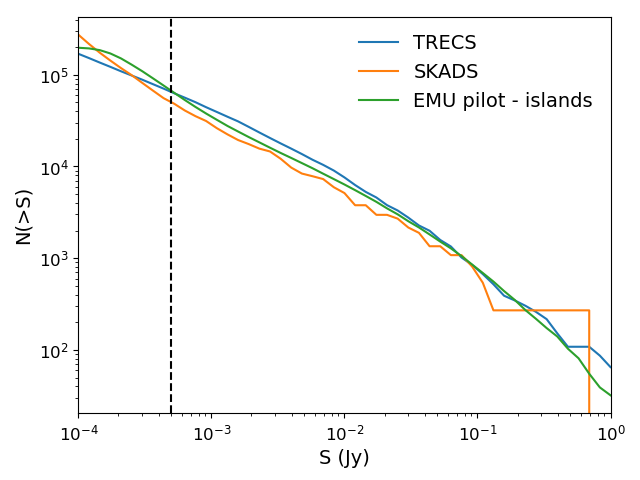}
    \caption{The number of sources in the EMU-PS with a flux density greater than some limit ($S$), as a function of that limit, compared to scaled predictions from the SKADS and T-RECS simulated catalogues. The dashed black vertical line gives the 500$\mu$Jy limit we assume for the clustering analysis presented in this paper. There is some discretisation of the prediction for the larger flux density limits, due to a scaling of some small integer value for the original prediction that was made for a much smaller value.}
    \label{fig:flux_distribution_emu_pilot}
    \end{center}
\end{figure}

The robustness of this approach to generate the randoms is checked by  comparing the fluxes of the simulated catalogues to the island catalogue of the EMU-PS. We selected both AGN and SFG galaxies from SKADS and T-RECS \citep{bonaldi19} simulations. In Figure \ref{fig:flux_distribution_emu_pilot} we compare the number counts for different flux density cuts $N(>S)$ between the EMU-PS island catalogue and the number counts from both simulated radio catalogues at $1$ GHz.

\begin{figure}
    \begin{center}
    \includegraphics[width=8cm]{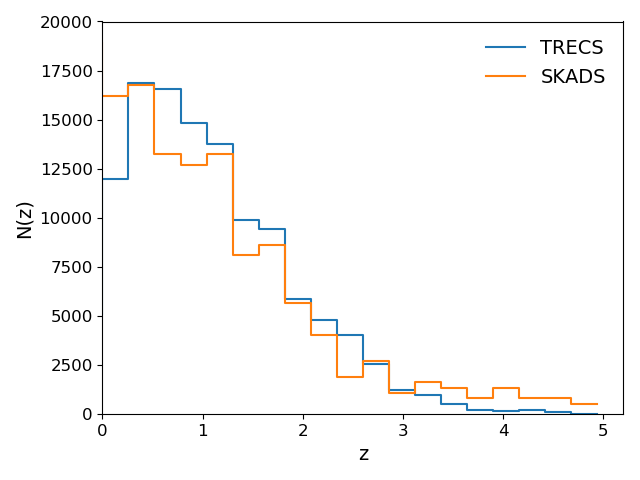}
    \caption{The predicted number of sources in the EMU pilot survey as a function of redshift, generated by scaling the predictions from the SKADS and T-RECS simulated catalogues. This assumes a flux density limit of 500$\mu$Jy}
    \label{fig:redshift_distribution_emu_pilot}
    \end{center}
\end{figure}

In order to calculate the expected  $w_{\mathrm LS}(\theta)$ distribution, we must know the redshift distribution $N(z)$ of the sources. For the analysis here, the SKADS and T-RECS simulations are used for  $N(z)$ and are shown in Figure \ref{fig:redshift_distribution_emu_pilot}. There is good agreement in the redshift distribution between these  two catalogues, and this consistency indicates that we are accurately modelling $N(z)$ and choosing the SKADS catalogue should not introduce a significant error. Using the $N(z)$ distribution estimated from SKADS, we compute  the theoretical expectation for the clustering statistics at the flux density cut of 500 $\mu$Jy.

\begin{figure}
    \begin{center}
    \includegraphics[width=8cm]{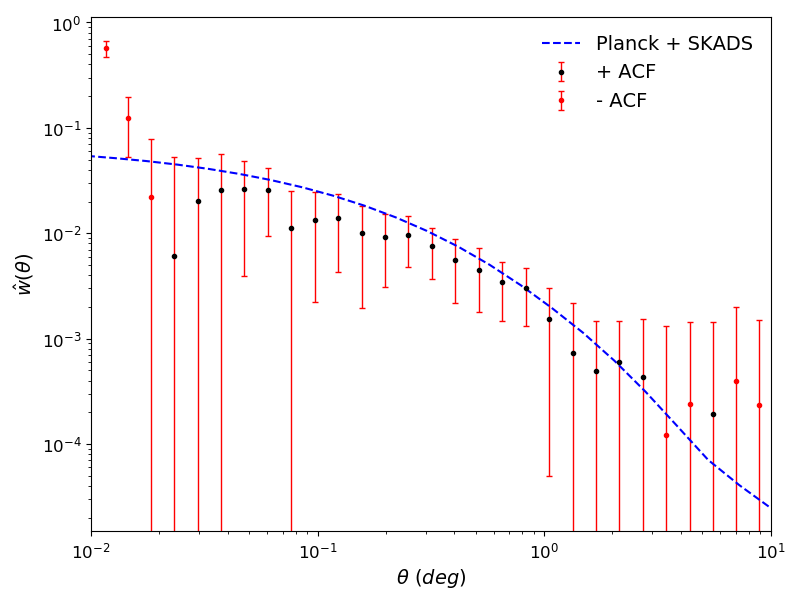}
    \caption{The measured angular correlation function \rev{(ACF)} $w(\theta)$ as a function of angular separation with one-sigma error bar computed from bootstrap re-sampling from 100 bootstraps. The correlation function is measured from the integrated flux-corrected EMU-PS island catalogue, using all sources above a flux density limit of 500$\mu$Jy. The blue curve is the theoretical prediction for the correlation function, assuming the Planck 2018 best fit cosmology and a SKADS model for the number distribution with redshift and the bias. No fitting of the cosmological or bias parameters was performed to change the prediction curve. \rev{As negative values cannot be shown on a log scale, in bins where the ACF becomes negative, we show (in red) the value of (-ACF) instead.}}
    \label{fig:cosmology_wtheta}
    \end{center}
\end{figure}

The measured  angular correlation function $w(\theta)$ is shown in Figure \ref{fig:cosmology_wtheta}, estimating the errors from boostrap resampling the data and random simulated data 100 times. We also show the predicted angular correlation function, assuming a cosmological model with values fixed \rev{at the values listed in Table} \ref{cosmoparams}, with a number distribution and bias model from SKADS. We show that the theoretical prediction, with no tuning of free parameters, is a reasonable fit to the data in the angular range $0.1\degr  < \theta < 10\degr$. There is somewhat of a discrepancy at small scales ($\theta < 0.1\degr$), which is probably generated by the multiple components that can be generated by the same radio galaxy, but which here are being treated as independent tracers of the cosmological density field.  A more complete analysis, including calibration of the effect of multi-component sources on the angular correlation function on small scales, is planned for future work.

\subsection{The excess of flat spectral index sources}

\label{flatspectrum}

To  investigate the skew towards flatter and inverted spectral indices shown in Figures \ref{spindex} and \ref{spindexh}, we compare the spectral index distributions for a sample of \rev{clearly} resolved and unresolved components from the EMU-PS.
The unresolved population  contains radio cores and  therefore includes flat-spectrum radio quasars \citep{Urry1995} and peaked-spectrum sources \citep{odea21}.
At the observing frequency of EMU-PS ($\nu \sim 900\,$MHz), the radio spectra of Gigahertz Peaked Spectrum sources tend to have shallow spectral index values as EMU-PS is observing close to their turnover frequency \citep{ODea1998}.
For this comparison we define each source to be resolved or unresolved using the component size after deconvolution from the beam, $\Psi$, considering components with $\Psi < 2$\arcsec\ to be unresolved and components with $\Psi > 20$\arcsec\ to be  resolved.

\begin{figure}
    \centering
    \includegraphics[width=\columnwidth]{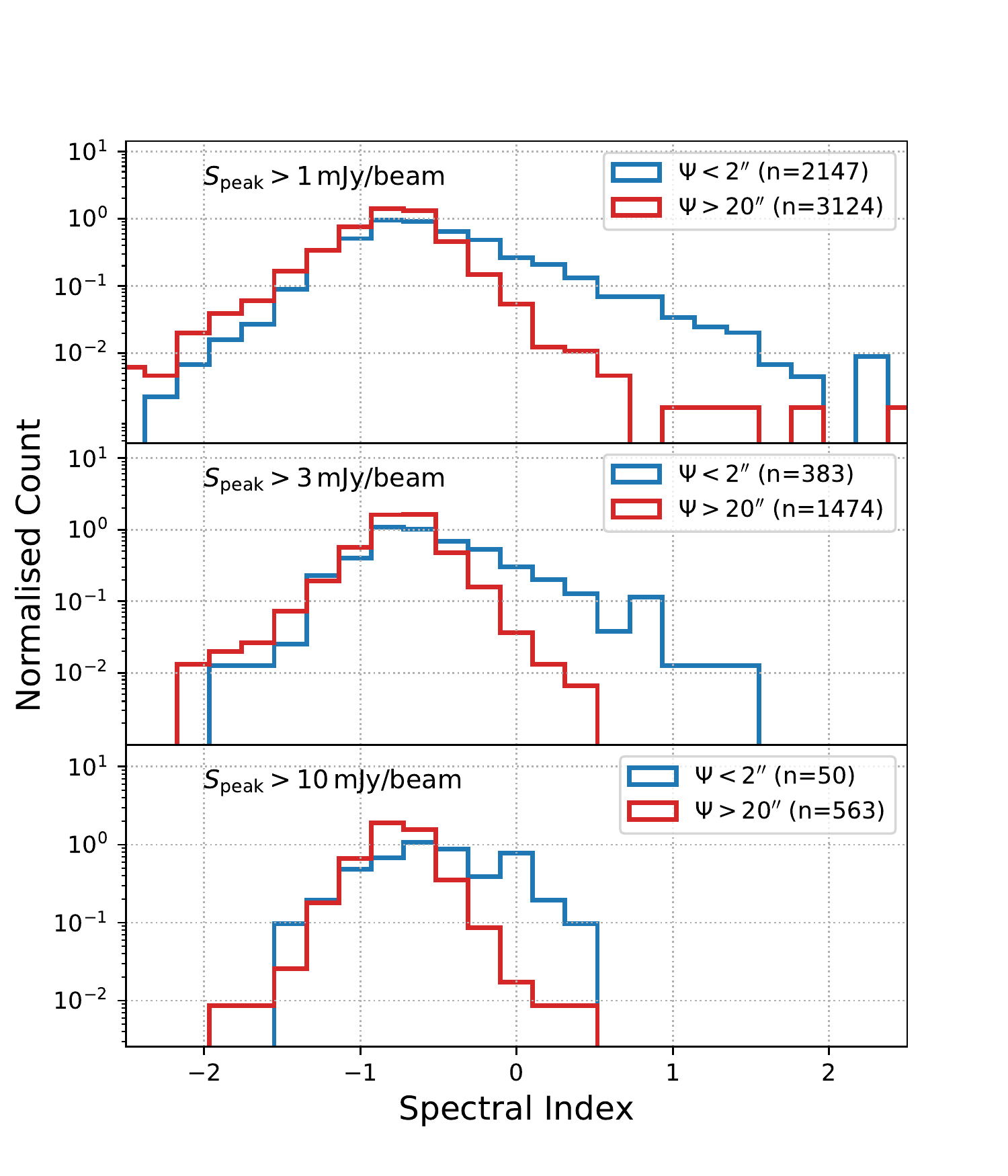}
    \caption{The spectral index distributions for unresolved ($\Psi < 2''$, blue) and resolved ($\Psi > 20''$, red) EMU components.
    The three panels show different minimum brightness levels, corresponding to $S_{\text{peak}} > 1\,$mJy/beam, $S_{\text{peak}} > 3\,$mJy/beam and $S_{\text{peak}} > 10\,$mJy/beam from top to bottom.
    The legend in each panel denotes the number of components contributing to each distribution shown. \rev{Each plot is normalised to the area under the curve.}}
    \label{fig:spidx_point_v_resolved}
\end{figure}

In Figure \ref{fig:spidx_point_v_resolved} we show the  spectral index distributions for the resolved and unresolved EMU components at three levels of minimum peak brightness: $S_{\text{peak}} > 1\,$mJy/beam, $S_{\text{peak}} > 3\,$mJy/beam and $S_{\text{peak}} > 10\,$mJy/beam.
Resolved components have symmetric distributions around a peak of $\alpha \sim -0.7$. 
However, while the spectral index distribution for the unresolved population is comparable to the resolved population at steep negative spectral indices ($\alpha < -0.7$), the distributions differ at flatter spectral indices.
At $\alpha > -0.7$ the unresolved population dominates over the resolved population at all three brightness levels.
Due to the large scatter in the spectral index distribution at sub-mJy levels (Figure \ref{spindex}), we have only performed this analysis on EMU components with $S_{\text{peak}}>1\,$mJy/beam.
Future EMU data, where potential issues in the spectral index calibration at lower signal to noise are better understood, will present the opportunity to study the spectral index distributions of fainter resolved and unresolved components.
Such an analysis, particularly for the full EMU survey, will enable tests of the potential flattening of the spectral index distribution for radio sources with $S \lesssim 0.5\,$ mJy \citep{Prandoni2006, Whittam2013}.

The different spectral index distributions of resolved and unresolved components we observe with the EMU-PS are consistent with what is seen with higher angular resolution and higher frequency observations. 
Recently, \citet{Gordon2021} demonstrated the $1.4 - 3\,$GHz spectral index distributions for point-like radio components are skewed to flatter values than well resolved components using observations from the Faint Images of the Radio Sky at Twenty cm survey \citep[FIRST; $\nu \sim 1.4\,$GHz,][]{Becker1995} and the Very Large Array Sky Survey \citep[VLASS, $\nu \sim 3\,$GHz,][]{Lacy2020}.
Further comparisons between EMU and radio observations in other bands, such as those from VLASS and the Australia Telescope Large Area Survey \citep[ATLAS, $1.4 - 2.3\,$GHz,][]{Zinn2012}, may help quantify the fractions of peaked-spectrum sources with different turnover frequencies, and this will be the focus of a follow-up work.

\section{Conclusion}
We have presented the first Pilot Survey of the Evolutionary Map of the Universe, using the Australian Square Kilometre Array Pathfinder (ASKAP) telescope. 

The resulting images reach an rms sensitivity of about 25--30 \ujybm\ rms at a spatial resolution of $\sim$ 11--18 arcsec, and result in a catalogue of $\sim$ 220,000  sources, of which $\sim$ 180,000 are compact. We have presented the catalogue of compact sources, together with optical and infrared cross-identifications and redshifts. We have also shown some preliminary science results, on both these compact sources and on diffuse sources,
which will be discussed in more detail in subsequent papers.

The results presented here testify  to the outstanding observational characteristics of ASKAP, including its  high survey speed and unprecedented sensitivity to low surface brightness emission.  Nevertheless, at the time of the Pilot Survey, several aspects of ASKAP correlator operation, calibration, and data processing were incomplete. We therefore expect future results from ASKAP, including the main EMU survey, to have even better sensitivity and dynamic range than the results presented in this paper.

\begin{acknowledgements}
\rev{We thank an anonymous referee for valuable feedback on an earlier iteration of this paper.} The Australian SKA Pathfinder is part of the Australia Telescope National Facility which is managed by CSIRO. Operation of ASKAP is funded by the Australian Government with support from the National Collaborative Research Infrastructure Strategy. Establishment of the Murchison Radio-astronomy Observatory was funded by the Australian Government and the Government of Western Australia. ASKAP uses advanced supercomputing resources at the Pawsey Supercomputing Centre. We acknowledge the Wajarri Yamatji people as the traditional owners of the Observatory site.
This work makes use of data products from the Wide-field Infrared Survey Explorer, which is a joint project of the University of California, Los Angeles, and the Jet Propulsion Laboratory/California Institute of Technology, funded by the National Aeronautics and Space Administration.
It also makes use of data from the European Space Agency (ESA) mission Gaia, and we acknowledge the institutions listed on \url{https://gea.esac.esa.int/archive/documentation/GEDR3/Miscellaneous/sec_acknowl/}
It also uses public archival data from the Dark Energy Survey (DES) and we acknowledge the institutions listed on \url{https://www.darkenergysurvey.org/the-des-project/data-access/}
This research has made use of the ``Aladin sky atlas'' developed at CDS, Strasbourg Observatory, France \citep{aladin}.
This research uses services or data provided by the Astro Data Lab at NSF's National Optical-Infrared Astronomy Research Laboratory. NOIRLab is operated by the Association of Universities for Research in Astronomy (AURA), Inc. under a cooperative agreement with the National Science Foundation.
Partial support for LR comes from US National Science Foundation Grant AST 17-14205 to the University of Minnesota.
The National Radio Astronomy Observatory is a facility of the National Science Foundation operated under cooperative agreement by Associated Universities, Inc.
CLH acknowledges support from the Leverhulme Trust through an Early Career Research Fellowship. IP acknowledges support from CSIRO under its Distinguished Research Visitor Programme, and from INAF through the SKA/CTA PRIN “FORECaST” and the PRIN MAIN STREAM “SAuROS” projects.
MJJ acknowledges support from the National Research Foundation of Korea under the program nos. 2017R1A2B2004644 and 2017R1A4A1015178.
\rev{CJR acknowledges financial support from the ERC Starting Grant `DRANOEL', number 714245. HA benefited from grant CIIC 174/2021 of Universidad de Guanajuato.}

\end{acknowledgements}

\bibliographystyle{pasa-mnras}
\bibliography{pilot}
\clearpage

\end{document}